\documentclass[pdflatex,sn-mathphys-num]{sn-jnl}

\usepackage{graphicx}
\usepackage{multirow}
\usepackage{amsmath,amssymb,amsfonts}
\usepackage{amsthm}
\usepackage{mathrsfs}
\usepackage[title]{appendix}
\usepackage{xcolor}
\usepackage{textcomp}
\usepackage{manyfoot}
\usepackage{booktabs}
\usepackage{algorithm}
\usepackage{algorithmicx}
\usepackage{algpseudocode}
\usepackage{listings}
\usepackage{bbold}

\begin{document}

\title[Article Title]{Ultrafast Non-Hermitian Skin Effect}

\author*[1]{\fnm{Barbara} \sur{Schneider}}\email{bschnei@phys.ethz.ch}\equalcont{These authors contributed equally to this work.}

\author*[1]{\fnm{Alexander} \sur{Dikopoltsev}}\email{adikopoltsev@phys.ethz.ch}\equalcont{These authors contributed equally to this work.}

\author[2]{\fnm{Markus} \sur{Bestler}}

\author[1]{\fnm{Philipp} \sur{T\"aschler}}

\author[1]{\fnm{Mattias} \sur{Beck}}

\author[3]{\fnm{David} \sur{Burghoff}}

\author[2]{\fnm{Oded} \sur{Zilberberg}}

\author*[1]{\fnm{J\'erome} \sur{Faist}}\email{jerome.faist@phys.ethz.ch}

\affil*[1]{\orgdiv{Institute for Quantum Electronics}, \orgname{ETH Zürich}, \orgaddress{\street{Auguste-Piccard-Hof 1}, \city{Z\"urich}, \postcode{8093}, \country{Switzerland}}}

\affil[2]{\orgdiv{Department of Physics}, \orgname{Konstanz University}, \orgaddress{\street{Universitätsstraße 10}, \city{Konstanz}, \postcode{78464}, \country{Germany}}}

\affil[3]{\orgdiv{Department of Electrical and Computer Engineering, Cockrell School of Engineering}, \orgname{The University of Texas at Austin}, \orgaddress{\street{301 E Dean Keeton St}, \city{Austin}, \postcode{78712}, \state{Texas}, \country{USA}}}

\abstract{Topological phases of matter commonly feature protected states at their boundaries. Transferring this protection to time-metamaterials is extremely challenging, as it requires the generation of an abrupt interface between two topologically distinct bulks. Here, we realize and measure an ultrafast topological non-Hermitian skin mode bound to an interface circulating within the cavity of a fast-gain semiconductor laser. The nonlinear stationary state generated in such devices features a jump in the instantaneous frequency. We show that this discontinuity gives rise to a topological interface for the field fluctuations in the system. Using direct intensity sampling, we experimentally measure the skin modes and their positioning at the frequency jump of the stationary state. Analysis of these isolated modes reveals an ultrashort full-width at half-maximum of 583 $\pm$ 16 fs. Furthermore, we show that we can tune the shape and relative timing shift of the skin modes via external bias modulation. Finally, both numerical and experimental analysis of the noise in the system reveal that field fluctuations are funneled into the topological interface. Our findings reveal a new way to generate topologically protected states of light in time, which paves the way for novel time-varying physics as well as metrological applications.}

\keywords{non-Hermitian topology, skin effect, spatiotemporal topology, quantum cascade lasers}

\maketitle

\section{Introduction}\label{sec1}
The study of topology dates back to the questions of whether the world is flat or round~\cite{Nakahara2018GeometryPhysics}. Answering these questions involves averaging over a local geometrical measure to determine the global topological structure.
In recent decades, band topology in bulk materials has brought topological phases of matter to the forefront of research~\cite{Bernevig2013TopologicalSuperconductors,Hasan2010ColloquiumInsulators,qi2011topological}. The quantization of observables in these phases reflects their nonlocal topological nature and underpins their robustness to local disorder. Moreover, when the topological order of a material is disrupted at its boundary, often, the bulk topology manifests itself as topologically protected boundary states in what is known as the bulk-boundary correspondence. The implications of topology in materials are extensively studied, with prominent examples including the quantum Hall effect with chiral edge transport~\cite{QuantumHallEffectBook}, topological insulators with boundary Dirac dispersion~\cite{Haldane1988ParityAnomaly,Hsieh2008DiracInsulator} or Majorana bound modes~\cite{Liang2008SuperconductingProximity}, and topological semimetals with Fermi arcs at their surface~\cite{Xu2015FermiArcs}. 

As the topology of materials is related to their crystalline structure, metamaterials with engineered band structures have emerged as fruitful media to demonstrate topological effects~\cite{Wang2008OneWayEdgeModes, Khanikaev2013PhotonicTopoInsu,Rechtsman2013FloquetTopoInsu,Hafezi2013ImagingEdgeState,Susstrunk2015MechanicalTopoInsu}. Such metamaterials are used in the field of topological photonics, where dielectrics are structured to generate an intricate optical dispersion with nontrivial band topology, and corresponding robust boundary states~\cite{Ozawa2019TopologicalPhotonics,Price2022RoadmapPhotonics,Lu2014TopoPhotonics}. The band structure of photonic systems is not filled with electrons up to a Fermi level, as in solids, but instead is probed by injected (bosonic) light. Thus, direct probing of topological boundary modes in a variety of photonic systems has been demonstrated, including topological pumps~\cite{Kraus2012QuasicrystalPumping,Zilberberg20184DQuantumHallPumping}, spin-Hall and valley-Hall effects~\cite{Hafezi2013ImagingEdgeState,Khanikaev2013PhotonicTopoInsu}, Floquet~\cite{Rechtsman2013FloquetTopoInsu} and square root topological insulators~\cite{Kremer202SquareRoot}, as well as high-order topology~\cite{Zilberberg20184DQuantumHallPumping,Noh2018MidGapModes}. Furthermore, such topological boundary phenomena is now harnessed for applications as protected communication channels for on-chip quantum operations~\cite{tambasco2018quantum}. 

In the same way that light is injected into the structured dielectric, it also leaks out of it. This engenders a driven-dissipative setting that has led to the study of non-Hermitian topology in open systems~\cite{Yan2023AdvancesPhotonics}. Commonly, this can be done by supplementing the effects of gain and loss to standard Hermitian topology models~\cite{Lieu2018TopologicalModel, Zeuner2015ObservationSystem,Hu2011AbsenceHamiltonians, Weimann2017TopologicallyCrystals}, where the environment solely embellishes the already existing topological features. However, another approach to open system topology involves novel topological effects that manifest solely due to the non-Hermitian properties of the system~\cite{Yao2018EdgeStatesInNHSystems,Su2021DirectSystem,Zheng2024DynamicDimensions,Li2024ObservationEffects,Fritzsche2024Paritytime-symmetricInsulator,Zhao2025Two-dimensionalGas}. Such non-Hermitian topological phases are classified with analogous indices to the Hermitian case, e.g., the spectral winding number~\cite{Okuma2019TopologicalNHSE}. The most prominent example of such models is the Hatano-Nelson model that describes a system in which energy is preferentially amplified in a specific direction of propagation~\cite{Hatano1996LocalizationMechanics}. More recently, it was established that non-Hermitian topology can manifest in a continuum bulk model with purely complex momentum-shifted particle-hole coupling~\cite{bestlerModel}. Interestingly, the eigenstates in such a bulk system (with periodic boundary conditions) are all extended, but when the boundaries are open, numerous modes become localized at the boundary in what is known as the non-Hermitian skin effect~\cite{Yao2018EdgeStatesInNHSystems,Wanjura2020TopologicalArrays,Cao2021DiffusiveFunneling, Zhao2025Two-dimensionalGas, Su2021DirectSystem,Sun2024PhotonicEffect,Okuma2020TopologicalEffects}. The directional flow induced by the unbalanced gain and loss effectively funnels all energy towards the non-Hermitian topological boundary of the system, which was demonstrated in time-bin encoded synthetic photonic lattices~\cite{Weidemann2020TopologicalLight,Leefmans2024TopologicalLaser}, and mechanical resonator chains~\cite{Wang2023Non-HermitianMetamaterials}, suggesting new locking mechanisms of light~\cite{Leefmans2024TopologicalLaser}.

Many-body interactions can give rise to exotic phenomena in topological Hermitian systems~\cite{Laughlin1983FQHE,Jain2009BookCompositeFermions,Clark2020LaughlinStatesFromLight,Stromer1999FractionalQuantumHall}, and their influence on non-Hermitian photonic topological models is a topic of growing contemporary interest.
In the driven-dissipative setting, we commonly encounter open Bose-Hubbard physics with weak photon-photon interactions (Kerr nonlinearity)~\cite{Vicentini2018DrivenBoseHubbard,Carusotto2013QuantumFluids}. When the interplay between drive, dissipation, and nonlinearity leads to system stabilization, the semiclassical stationary states can have topological properties. Examples include robust discrete solitons~\cite{Szameit2024DiscretePhotonics}, self-induced topological protection~\cite{Liu2021Gain-inducedInteractions}, or even topological lasing~\cite{St-Jean2017LasingLattice,Liu2021Gain-inducedInteractions,Bahari2017NonreciprocalGeometries,Harari2018TopologicalTheory,Bandres2018TopologicalExperiments,Amelio2020TheoryLasers,Zeng2020ElectricallyModes,Dikopoltsev2021TopologicalArray}. 
Nonlinearity also acts as a tool to introduce time dependence to topological systems and to generate new topological phases~\cite{Lustig2018TopologicalCrystals}. For example, introducing soliton generation mechanisms can lead to topologically mode-locked sources~\cite{Mittal2021TopologicalSolitons,Flower2024ObservationCombs, Leefmans2024TopologicalLaser}. Other time-varying systems suggest energy-momentum band gaps~\cite{Moussa2023ObservationInterfaces, Mendonca2002TimeConcepts,Sharabi2022SpatiotemporalCrystals} and demonstrate space-time topology~\cite{Feis2024Spacetime-topologicalEvents,Ren2025ObservationLattice}. In principle, spatiotemporal stationary states of nonlinear driven-dissipative systems~\cite{Shandilya2025,Wang2018} can give rise to excitations with underlying nontrivial indices~\cite{Lado2019HarperHeisenberg}. However, it is extremely challenging to produce sufficient changes in time, much faster than the time scales of electronics~\cite{Lustig2023Time-refractionModulation}, to impact the topological properties of light. 

In this work, we demonstrate that the extended nonlinear state of a fast-gain laser produces a spatiotemporal non-Hermitian topological interface. The state of our fast-gain laser supports Bogoliubov excitation modes on top of a propagating kinked phase profile, which forms an ultrafast topological skin-effect interface between continuous bulks for these excitations. We explain this formation by mapping the fluctuation dynamics to a complex momentum-shifted particle-hole coupling. We measure the excitation using a direct asynchronous sampling method and find that it is locked to the spatiotemporal topological interface. Moreover, the intracycle timing of the interface can be altered by detuning the modulation of the cavity. Interestingly, we show that the full width at half-maximum of the excitation can drop below that of the transform-limited pulse given by the spectrum of the underlying signal. Such a short duration of the time-interface and the topological states, below 1 ps, results from fast intraband electronic times. The intrinsic formation of the  temporal skin-effect paves the way to new topological phases, locked entities in time for precision measurements, and novel experimental approaches to time-switching of materials~\cite{Gunter2009Sub-cycleInteraction}. 

\section{Results}\label{sec2}
\begin{figure*}[htbp]
\centering
\includegraphics[width=\textwidth]{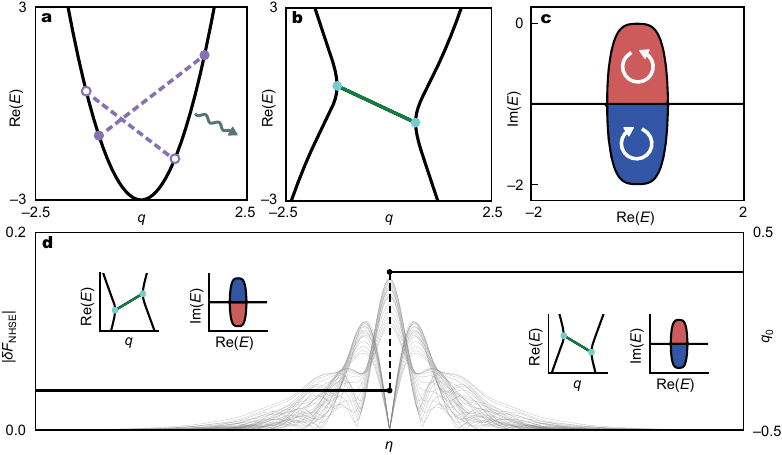}
\caption{\textbf{a}, Sketch of the model described in Eq.~\eqref{eq:MainHamiltonian} for $q_0=0.3$. Bosonic modes with parabolic dispersion (black line) experience parametric pair creation (annihilation) processes with an additional momentum shift $2q_0$. Example processes are illustrated by full (empty) purple circles connected via dashed lines. \textbf{b}, Dispersion relation of the system depicted with periodic boundary conditions for $m=1/4$, $\Gamma=1$ and $q_0=0.3$. The exceptional points are marked in light blue and the diabolical line is marked in green. \textbf{c}, Spectrum of the system with periodic boundary conditions (black lines) for the same parameters as \textbf{b}. Regions with spectral winding number $1$ ($-1$) are marked in red (blue). \textbf{d}, A topological interface arising through a change in the sign of $q_0$ in space $\eta$. Across the topological interface, the spectral winding numbers [cf.~\textbf{c}] flip sign, marking two topologically distinct regions. We show all numerically calculated non-Hermitian skin modes that are localized at this interface (grey lines).}
\label{fig:Bestlermodel}
\end{figure*}

We start by succinctly reviewing how non-Hermitian skin modes arise via a purely complex momentum-shifted parametric process, see Fig.~\ref{fig:Bestlermodel}. We consider the model~\cite{bestlerModel}
\begin{align}
    {\displaystyle{H=\int dq \left[(\frac{q^2}{2m}-i\Gamma) a_{q}^\dagger a_{q}^{\phantom \dagger} - \left(i\Gamma \,a_{q}^\dagger a_{-q-2q_0}^{ \dagger} + h.c.\right) \right]\, },}
    \label{eq:MainHamiltonian}
\end{align}
where $m$ is the effective mass which acts as a dispersion, and $\Gamma$ is the strength of the parametric process between two modes shifted in momentum by a wavenumber $2q_0$. Here, $\Gamma$ is  also the rate of an overall dissipation strength that stabilizes the system from parametric instabilities, see Fig.~\ref{fig:Bestlermodel}a. This model can be mapped to level attraction between particle-like Bogoliubov modes with positive dispersion and its shifted conjugate hole-like modes with negative dispersion~\cite{bestlerModel}. When the parametric process is strong enough ($|\Gamma| > q_0^2$), the two bands coalesce to the spectrum shown in Fig.~\ref{fig:Bestlermodel}b. This creates a tilted diabolical line that is terminated by two exceptional points. The diabolical line leads to the formation of two loops in the complex spectrum of the system, see Fig.~\ref{fig:Bestlermodel}c. Crucially, the (non-Hermitian) local dissipation ensures that every mode still has a finite lifetime, manifesting in eigenenergies with a negative imaginary part.

Due to the non-Hermitian bulk-boundary correspondence, the spectral structure of the system is intimately tied to the emergence of edge modes when interfaced with a topologically distinct phase. The model~\eqref{eq:MainHamiltonian} can be characterized by a spectral winding number $W(E)$ (see SI for details). A positive (negative) value of $W(E)$ at a given complex energy $E$ signals a skin mode localized at the right (left) boundary. For the two spectral arcs, we find $W(E)=\pm1$ (see Fig.~\ref{fig:Bestlermodel}c), indicating the presence of topological skin modes at both edges.
This prediction is confirmed by interfacing the system with an identical copy where the sign of $q_0$ is reversed across the boundary. As shown in Fig.~\ref{fig:Bestlermodel}d, skin modes emerge at the interface, consistent with a reversal of the winding number. The boundary thereby functions as a domain wall between topologically distinct bulk phases, enabling the localization of modes from both sides \cite{Jackiw1997CurrentSoliton, TriangularWormholes2019Sayan}.
In the next section, we demonstrate that such interfaces naturally arise in fast-gain lasers, where they support non-Hermitian spatiotemporal topological skin modes.

\subsection{The Extendon State and the Ultrafast Topological Boundary}\label{subsec21}

\begin{figure*}[htbp]
\centering
\includegraphics[width=\textwidth]{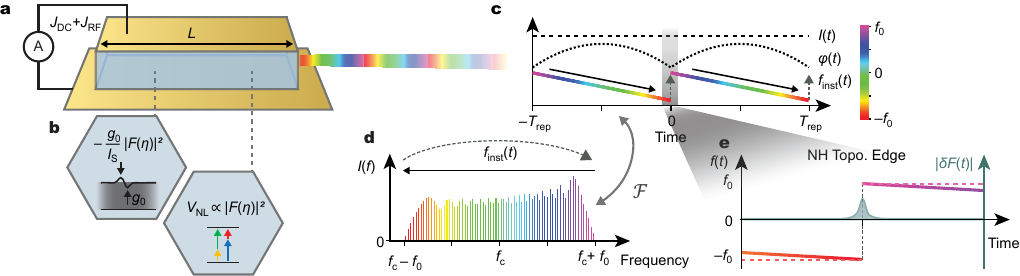}
\caption{Extendons in fast-gain lasers as a platform for ultrafast topology. \textbf{a}, Semiconductor laser biased with a constant bias current $J_\text{DC}$ superimposed with a modulated bias current $J_\text{RF}$ emitting broadband periodic light. \textbf{b}, Internal nonlinear processes induced by the gain recovery. (left) Intensity saturation that keeps the intensity constant and suppresses fluctuations on ultrashort timescales. (right) Four-wave mixing through spatial hole burning, which is dependent on the phase. \textbf{c}, The nonlinear processes in \textbf{b} generate an extendon, quasi-constant intensity periodic signal with a varying phase. The phase $\phi(t)$ contains a kink that is translated to a jump in the instantaneous frequency $f_\text{inst}(t)$. \textbf{d}, the frequency spectrum of the extendon with arrows indicating the frequency sweep across the spectrum during each repetition period (solid, black) and the frequency reset between the edges of the spectrum between the periods (dashed, grey). \textbf{e}, in the vicinity of the frequency jump, frequencies from both sides are approximated to constant $\pm f_0$, experiencing a non-Hermitian topological interface, as described in Fig.~\ref{fig:Bestlermodel}. This interface supports non-Hermitian skin modes of the field fluctuations (pale green).}
\label{fig:extendon}
\end{figure*}

To generate and probe non-Hermitian skin modes, we employ mode-locking in a Fabry-Pérot semiconductor laser with ultrafast gain recovery  enabled by intraband scattering processes~\cite{Khurgin2014, Opacak2019b, Burghoff2020d, Heckelmann2023QuantumLaser, Taschler2023h}, see ~\ref{fig:extendon}a.
The ultrafast recovery time implies that fluctuations in the intracavity power are immediately suppressed~\cite{Dikopoltsev2025TheLaser}, see Fig.~\ref{fig:extendon}b.
 Using a mean-field description that unfolds the forward and backward propagating fields into a $2L$ effective roundtrip, the dynamics of the normalized field $F$ in the co-moving frame reduces to a generalized nonlinear Schrödinger equation (NLSE)~\cite{Burghoff2020d, Heckelmann2023QuantumLaser,Dikopoltsev2025CollectiveDimensions}:
\begin{equation}
    i\frac{\partial F}{\partial T} = -\frac{k''}{2} \frac{\partial^2F}{\partial t^2} + V(F)F+ig(1-\left|F\right|^2/P_\text{s})F-i\alpha F,
    \label{eq:NLSE}
\end{equation}
where $t$ is the fast intracycle time coordinate and $T$ the slow-time evolution of the system. The parameter $k''$ represents the group velocity dispersion of the cavity, and the term $ig\left(1 - |F|^2/P_s\right)F$ captures the gain saturation, where $g$  is the linear gain, $P_s$ is the saturation power, and $\alpha$ are the losses. This term enforces quasi-constant intensity over the full cycle and enables mode-locking~\cite{Dikopoltsev2025TheLaser}. The nonlinear potential $V(F)$ describes the mechanism responsible for mode proliferation. In systems with fast gain recovery, a $V(F)$ that depends only on $|F|^2$, as typically used in soliton models, is insufficient, since it produces only a global phase shift.
Instead, in Fabry-Pérot cavities with slow carrier diffusion, $V(F)$ depends explicitly on the phase $\phi$ of $F$ and takes the form $V(F) \propto \phi |F|^2$, see Fig.~\ref{fig:extendon}c. This structure is known as an \textit{extendon}, in analogy to solitons, which are localized solutions~\cite{Burghoff2020d, Humbard2022AnalyticalStates}. Figure~\ref{fig:extendon}d shows the optical mode spectrum of an extendon, with arrows indicating the frequency sweep over a single cycle. The state exhibits a discrete jump in instantaneous frequency at each repetition period, which we identify as an ultrafast boundary between two distinct spatiotemporal bulk phases with opposite topological winding numbers, see  Fig.~\ref{fig:extendon}e.

We now develop the Bogoliubov expansion of such a system to find the bulk properties of the field fluctuations $\delta{F}=F-F_0$ (see derivation in the SI), which are  the variation from an extendon steady state $F_0$. The dynamics of the fluctuations~\cite{Chiocchetta2013Non-equilibriumDimensions} are  described by 
\begin{align} 
\label{eq:FluctuationEOM} i\frac{\partial }{\partial T} \delta F=-D \frac{\partial^2}{\partial \eta^2}\delta F-i \Gamma \delta F-i\Gamma e^{2i \phi(\eta)}\delta F^*, 
\end{align} 
where $\eta=Kz-\Omega t\in[0,2\pi]$ is a co-rotating coordinate with $K=2\pi/L$ and $\Omega=Kc$, $D$ is the dispersion, $\Gamma=g-\alpha$ is the non-Hermitian coupling strength, and $\phi(\eta)$ is the phase of the steady-state field. We observe that fluctuations are coupled to their conjugate with a factor that depends on $\phi(\eta)$. We note that this type of coupling corresponds to squeezing and anti-squeezing terms (see SI).  In the states that we study, on each side of the frequency jump, we assume $\phi(\eta)$ to be linearly dependent on the coordinate $\eta$ with $\frac{d\phi(\eta)}{d\eta}=\pm |q_0|$ (Fig.~\ref{fig:extendon}e). By explicitly expanding the field $\delta F=(1/\sqrt{2\pi})\int dk e^{iq\eta}\hat{a}_q$ on each side of the discontinuity, our linearized fluctuation model is directly translated to the non-Hermitian topological interface from Eq.~\eqref{eq:MainHamiltonian} (see SI). As such, we establish that the extendon supports a non-Hermitian topological boundary that is co-rotating at the speed of light with the underlying state. 

\subsection{Measurement}\label{subsec22}

\begin{figure*}[htbp]
\centering
\includegraphics[width=\textwidth]{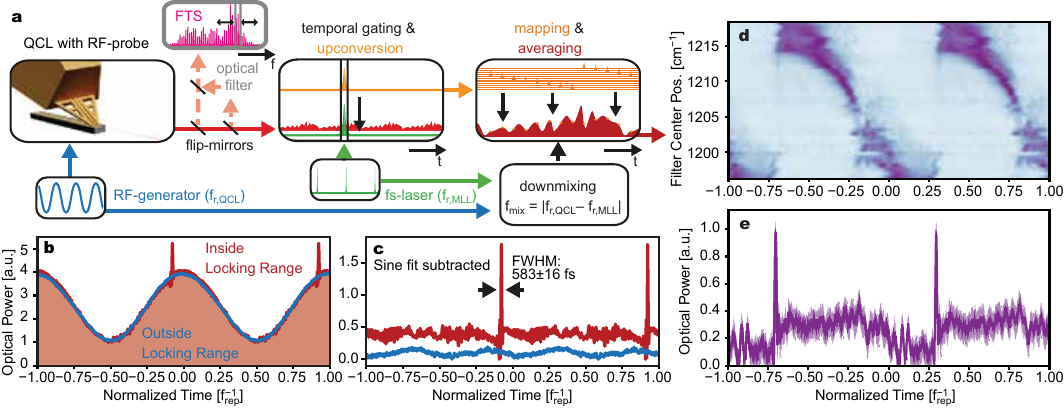}
\caption{Measurement of the spatiotemporal topological boundary states using asynchronous upconversion sampling (ASUPS). \textbf{a}, Conceptual schematic of the measurement method, using external RF-modulation to stabilize the repetition rate of the output and as a measurement reference. The modulation frequency, $f_\text{r,QCL}$, is 11 GHz, the repetition rate of the pulsed probe laser is, $f_\text{r,MLL}$ 90 MHz \textbf{b}, Intensity of the signal when the repetition rate is locked (red) and when it is not locked (blue) to the external modulation at a nominal output power of the RF-generator of 20 dBm. \textbf{c}, The same signal after subtracting a constant offset and a sinusoid, corresponding to steady-state power of the extendon and the gain modulation caused by the bias modulation. \textbf{d}, The measured intensity as a function of the position of an optical filter placed between the laser and the ASUPS measurement setup. \textbf{e}, The signal measured when opening the slit.}
\label{fig:meas}
\end{figure*}

To measure the non-Hermitian skin effect in the time domain, we use extendons that form spontaneously in Fabry-P\'erot mid-infrared quantum cascade lasers~\cite{Faist1994}, with an ultrafast gain recovery time of around $0.5$ ps. As the spectral span of such extendons in our system is on the order of 1 THz, the model implies that the corresponding boundary states will have a duration of the order of 1 ps~\cite{bestlerModel}. Alas, the use of typical methods of time-domain signal recovery by phase-sampling in the frequency domain~\cite{Han2020a} is exposed to the detrimental effects of phase-noise, which masks narrow features in the time-domain. Instead, we use  direct sampling with the asynchronous upconversion sampling (ASUPS) method~\cite{Taschler2023h}. This method uses an external femtosecond laser, with a full width at half-maximum (FWHM) of approximately 100 fs, that asynchronously samples the mid-infrared output of our laser via sum-frequency generation. The experimental setup used for all measurements is shown schematically in Fig.~\ref{fig:meas}a. 

In Fig.~\ref{fig:meas}b, we show the output of an externally modulated QCL, within and outside its locking range. The modulation is primarily performed to stabilize the repetition rate~\cite{Schneider2021c} and provide a clean reference. We compare the locked and unlocked traces to confirm that the sinusoidal modulation of the intensity is not a coherent effect, but rather the result of the gain modulation caused by the external modulation. To qualitatively isolate the features of the signal, the traces in Fig.~\ref{fig:meas}c display the same data as in Fig.~\ref{fig:meas}b, but after removing a sinusoidal fit from the data. We observe a spike in intensity when inside the locking range, which is predicted to be the non-Hermitian skin modes formed by the fluctuations and has a FWHM of 583 $\pm$ 16 fs. 

To verify that the very nature of the topological excitation is defined by its strong confinement to the jump in instantaneous frequency, we use an optical filter before the ASUPS setup to add spectral resolution to the time-domain measurement. The results of a sweep in the filter position are shown in Fig.~\ref{fig:meas}d. The monotonic frequency chirp~\cite{Singleton2018a, Taschler2023h, Hillbrand2019} is reproduced, including the discontinuity between the edges of the spectral range, as depicted in Fig.~\ref{fig:extendon}c. The ASUPS trace with the filter fully open is shown in Fig.~\ref{fig:meas}e, where we have removed the DC intensity offset for clarity. It is evident in the experimental data that the excitation is located at the jump in instantaneous frequency. Supporting simulation data are provided in the SI.

\subsection{Confinement of the Topological Mode}\label{subsec23}

\begin{figure}[htbp]
\centering
\includegraphics[width=\textwidth]{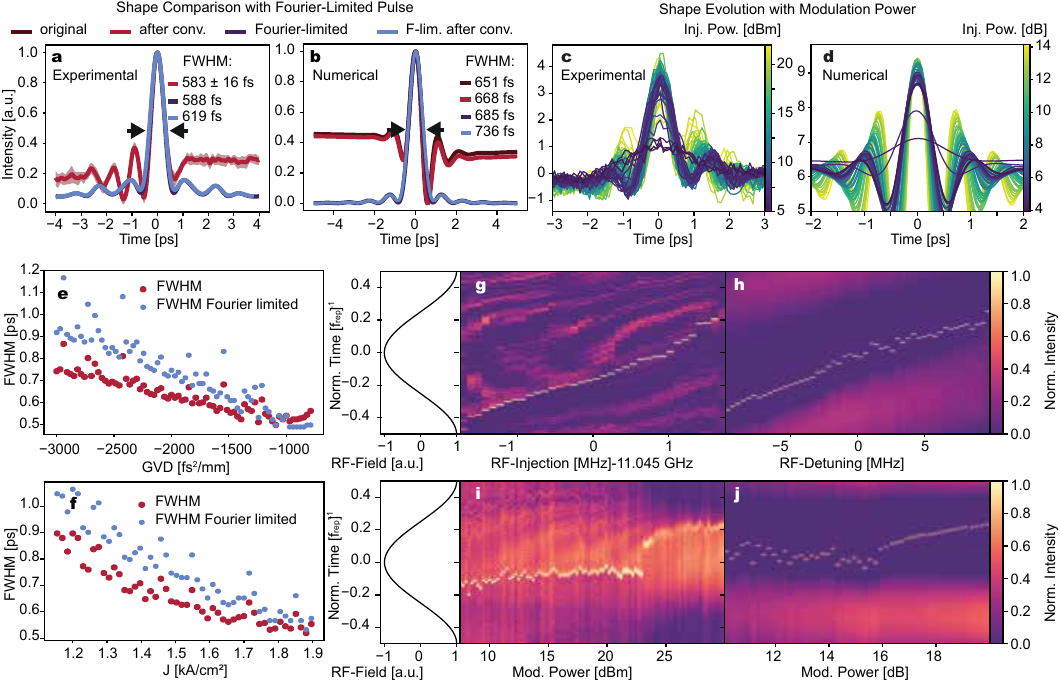}
\caption{Measured non-Hermitian skin modes and the effects of power and frequency of the modulation. \textbf{a}, The isolated skin modes (red), the transform-limited pulse calculated from the extendon spectrum (blue), the same pulse convolved with a pulse of the width of the ASUPS probe for comparability  (light blue). \textbf{b}, Numerical simulations of the states in \textbf{a}, with the convolution of the skin modes plotted using the same color coding as in \textbf{a}. The original modes without undergoing a convolution are shown in dark red. \textbf{c}, The experimental skin modes measured at different bias modulation powers and \textbf{d}, the corresponding simulated signal. \textbf{e}, The skin modes FWHM vs. the dispersion in numerical simulations. \textbf{f}, The skin modes FWHM upon changing the current density in the simulations. Both \textbf{e} and \textbf{f} show that the width of the topological state can drop below that of the transform limited signal. \textbf{g}, The intracavity intensity at different modulation frequencies, and \textbf{h}, the corresponding numerical simulations.\textbf{i}, The intracavity intensity  for different modulation power the corresponding simulations \textbf{j},. The relative position of the skin modes shifts with respect to the frequency and amplitude of the modulation. }
\label{fig:manip}
\end{figure}

To assess the confinement of the skin modes to the interface, we isolate the intensity spike and compare it to the transform-limited pulse based on the power spectrum. Moreover, we show the convolution of the transform-limited Gaussian pulse with a FWHM of 100 fs to account for any broadening caused by the finite width of the probe pulse. In Fig.~\ref{fig:manip}a, we observe that the measured skin modes intensity  has a FWHM of 583 $\pm$ 16 fs, which is narrower than both transform-limited calculated pulses, whose FWHMs are 588 fs and 619 fs. This means that the duration of the topological states drops below the bandwidth limit of the underlying signal. We studied this further in simulations by tuning the cavity dispersion and pump current density, as shown in Fig.~\ref{fig:manip}e and f. We found that the ratio between the widths of the non-Hermitian skin modes and the transform-limited pulse of the underlying extendon spectrum can drop down to 0.8, at least for this range of parameters. Such considerations show potential for optimizing the topological excitation-width for applications such as  time-domain probes.

The duration of the non-Hermitian skin modes is inversely proportional to the size of the instantaneous frequency jump and can be impacted by modulation. The current modulation translates to phase-modulation through the linewidth enhancement factor in such devices~\cite{Harder1983a, Aellen2006a, Opacak2021SpectrallyComb}, and contributes to spectral broadening. By changing the modulation power of the laser, we therefore expect a change in the frequency jump~\cite{Schneider2021c}. In Figs. \ref{fig:meas}c,d, we present the measured skin modes under an increase in modulation that widens the jump and therefore causes a stronger state confinement, with the corresponding numerical simulation, respectively. Indeed, we find that upon increasing the modulation power, the power of the non-Hermitian skin modes increases and its FWHM decreases.

Finally, we show that the relative timing of the ultrafast non-Hermitian skin-effect inside the cycle can also be manipulated. Here, we use frequency detuning and modulation power that shift the frequency jump within the cycle. This kind of shift of the extendon was theoretically predicted in earlier work in the absence of the phase-nonlinearity term $\propto\phi|E|^2$~\cite{Dikopoltsev2025TheLaser}, where the shift was $t_0 \sim \arcsin{\frac{\Delta L_c}{8\sqrt{M\beta}}}$. This amounts to an approximately linear relative shift close to the center of the locking range. Likewise, we find quasi-linear tuning of the skin modes position in time when we keep the modulation power constant and sweep the frequency across the locking range. In Figs. \ref{fig:manip}g,h, we show the intracavity intensity for a range of detuning frequencies, experimentally measured and numerically simulated, respectively. We find high qualitative agreement, showing that the skin modes changes position with the same character as predicted above. Similarly high agreement is found when keeping the frequency constant and sweeping the modulation power. In Fig.~\ref{fig:manip}i, we show  no significant change in position for lower injection powers, followed by an approximately $\frac{\pi}{2}$ phase-shift at sufficiently high modulation. The same can be found in the numerical simulation, where after a certain modulation depth, the non-Hermitian skin modes changes relative position, see Fig.~\ref{fig:manip}j. This clearly demonstrates that the time-locked pulse can be shifted and its location can be predicted, which is a key to various metrological applications.

\section{Fluctuation Funneling}\label{sec24}

\begin{figure}
    \centering
    \includegraphics[width=\textwidth]{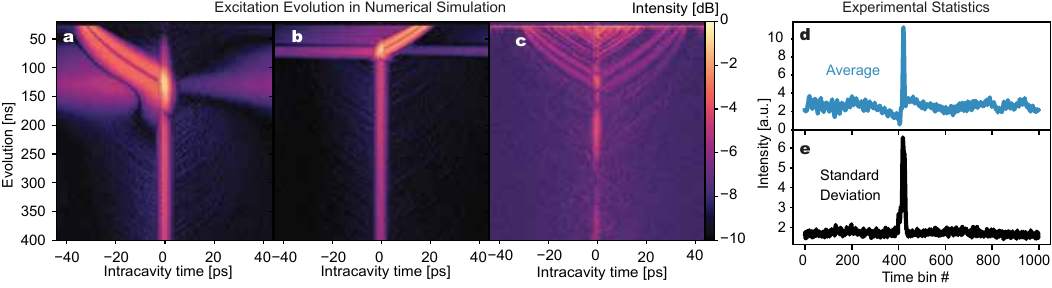}
    \caption{Time-domain funneling towards the interface. Local fluctuations at different times in the cycles, funneling towards the non-Hermitian topological interface. \textbf{a}, When before interface, \textbf{b}, when behind interface, and \textbf{c}, when uniformly excited at all times. \textbf{d}, Average intensity measured using ASUPS as a function of the time bins used for data processing. \textbf{e}, The standard deviation calculated from the same dataset.}
    \label{fig:funneling}
\end{figure}

In analogy to the fluctuation dynamics for the non-Hermitian skin effect in the spatial domain~\cite{Weidemann2020TopologicalLight}, we can study the noise properties in our lasing states and show both theoretically and experimentally that the fluctuations of the field are funneled towards the topological interface. Using Eq.~\eqref{eq:NLSE}, we calculate the evolution of local fluctuations with a well-defined frequency for multiple cavity roundtrips. We observe that any fluctuations will propagate in the co-rotating frame towards the temporal interface, here at $t=0$, where energy accumulates and acts as a source for the topologically locked state, as shown in Fig.~\ref{fig:funneling}a for excitations on either side of the interface. We perform a very similar numerical calculation in Fig.~\ref{fig:funneling}b, but initiate the system with a flash of white noise, and observe that the long-lived modes are the ones that flow towards the interface, meaning that excess noise will naturally accumulate at the boundary. We have observed this accumulated excess noise experimentally by performing statistics on the direct intensity measurements performed with ASUPS: in Fig.~\ref{fig:funneling}c, we show the standard deviation of the intensity within the period, and the accumulation of noise near the discontinuity point of the phase derivative.
\section{Discussion}\label{sec12}

In this work, we have directly measured ultrafast topological non-Hermitian skin modes supported by the fast intraband recovery times in semiconductor lasers. These states are shown to be confined to the discontinuity in the instantaneous frequency, which is inherent to the extendons formed in fast-gain laser. We have shown that this state is ultrashort with controllable width that can drop below the transform-limited pulse width dictated by the spectrum of the underlying extendon. We have demonstrated the controllability over the topological interface, and changed its relative position and size using RF-injection, thereby shifting and shaping the topologically bound state. Our comparison with numerical simulations based on a mean-field theory demonstrates the strongly predictive nature for this state. Finally, we showed that noise in the system follows the prediction on the dispersion of the fluctuation and its topological boundary. Our work drives progress in both the field of time-dependent phenomena and topological systems with states that can potentially be used in novel meteorological and communication applications.

\section{Methods}\label{sec11}

For the experimental measurements we use a 4 mm long RF-enhanced~\cite{Kapsalidis2021e, Schneider2021c} QCL biased at 1.23 A, which we modulate close to its free-running beatnote frequency around 11.045 GHz to stabilize it in order to enable measurement of any narrow features in the time-domain. The RF powers used in the experiments are the nominal values set for the output of our RF-generator in the idealized case of a 50 $\Omega$ load.

\backmatter

\bmhead{Supplementary information}

If your article has accompanying supplementary file/s please state so here. 

\bmhead{Acknowledgements}

We are grateful for fruitful discussions with J.~del Pino and I.~Carusotto. We acknowledge funding from the European Research Council (ERC) under the European Union’s Horizon 2020 research and innovation programme (Grant agreement No. 820419) and from the Swiss Innovation Agency Innosuisse (Grant agreement No. 2155008433). ETH Fellowship program: (22-1 FEL-46) (to AD). The Konstanz team acknowledges funding from the Deutsche Forschungsgemeinschaft (DFG) via project numbers 449653034; 425217212; 521530974; 545605411;  and through SFB1432, as well as from the Swiss National Science Foundation (SNSF) through the Sinergia Grant No.~CRSII5\_206008/1.

\begin{appendices}

\section{Spectral Winding Number}
For calculating the spectral winding number of the bulk model, we follow Ref.~\cite{bestlerModel}.
We consider the model in Eq.~\eqref{eq:MainHamiltonian} describing the complex momentum-shifted parametric process with local damping
\begin{align}
    {\displaystyle{H=\int dq \left[(\frac{q^2}{2m}-i\Gamma) a_{q}^\dagger a_{q}^{\phantom \dagger} - \left(i\Gamma \,a_{q}^\dagger a_{-q-2q_0}^{ \dagger} + h.c.\right) \right]\, }.}
\end{align}
It is instructive to switch to a momentum-shifted frame $q'=q+q_0$. We then rewrite $H$ using a bosonic Nambu spinor with shifted momentum $\mathbf{a}_q'=(\hat a_{q'-q_0}, \hat a_{-q'-q_0}^\dagger)$ to $H=\int dq' \mathbf{a}_{q'}^\dagger (\mathbf{H}_{q'}-i\Gamma\mathbb{1}) \mathbf{a}_{q'}$ with a $2\times2$ Hamiltonian density $\mathbf{H}_{q'}$. The spectrum of the system is given by the eigenvalues of the corresponding dynamical matrix \cite{Xiao2009TransformationTheory}
\begin{equation}
    \mathbf{D}_{q'}\equiv\sigma_z\mathbf{H}_{q'}-i\gamma\mathbb{1} =
    \begin{pmatrix}
        \frac{1}{4m}(k-k_0)^2-i\Gamma & -i\Gamma\\
        -i\Gamma & -\frac{1}{4m}(-k-k_0)^2-i\Gamma
    \end{pmatrix}\,.
    \label{eq:DynamicalMatrix}
\end{equation}
The spectral winding number for this bosonic continuum system then reads
\begin{align}
W(E) = \int_{-\infty}^{\infty} \frac{d q'}{2\pi i}\frac{d}{dq'} \log(\det(\mathbf{D}_{q'}(q')-E))\,,
\label{eq:WindingNumber}
\end{align}
where $E\in\mathbb{C}$ is a complex energy.

\section{Derivation of the Fluctuation Equation}
We develop the equation for the fluctuation dynamics in the system, Eq. \ref{eq:FluctuationEOM} , when it is in a steady state of an extendon from of $F_0=\sqrt{P_0}e^{i\phi(\eta)}$, where $P_0$ is the steady state power. Assuming the state is $F(\eta,T)$, the fluctuations are given by $\delta F(\eta,T)=F(\eta,T)-F_0(\eta)$. To describe the dynamics induced by the fast gain recovery times, we substitute this expression into Eq. \ref{eq:NLSE}  
\begin{align}
    \begin{aligned}
    i\frac{\partial}{\partial T}(\delta F+F_0) =& -\frac{k''}{2} \frac{\partial^2}{\partial t^2} (\delta F+F_0)+ V(\delta F+F_0)(\delta F+F_0)\\
    &+ig(1-\left|(\delta F+F_0)\right|^2/P_\text{s})(\delta F+F_0)-i\alpha (\delta F+F_0),
    \end{aligned}
\end{align}
This expression can be simplified by using that $V$ is slowly varying with $F$ so that $V(F)\approx V(F_0)$, and that $F_0$  is solving Eq. \ref{eq:NLSE}:
\begin{align}
    \begin{aligned}
    i\frac{\partial}{\partial T}\delta F =& -\frac{k''}{2} \frac{\partial^2}{\partial t^2} \delta F+ V(F_0)\delta F+i(g-\alpha)\delta F\\&+ig(-\left|(\delta F+F_0)\right|^2/P_\text{s})(\delta F+F_0)-ig(-\left|F_0\right|^2/P_\text{s})F_0,
    \end{aligned}
\end{align}
We use the first order perturbation, neglecting $O(\delta F^2)$, and the steady state saturation condition of balanced loss and gain, $g=\alpha +gP_0/P_s$., to find
\begin{align}
    \begin{aligned}
    i\frac{\partial}{\partial T}\delta F &= -\frac{k''}{2} \frac{\partial^2}{\partial t^2} \delta F+ V(F_0)\delta F+i(g-\alpha)\delta F-ig/P_\text{s}(F_0^2\delta F^*+2\left|F_0\right|^2\delta F)\\&= -\frac{k''}{2} \frac{\partial^2}{\partial t^2} \delta F+ V(F_0)\delta F-ig/P_\text{s}(F_0^2\delta F^*+\left|F_0\right|^2\delta F),
    \end{aligned}
\end{align}
We define $\Gamma=gP_0/P_s$, the dispersion coefficient in the coordinate $\eta$ is $D=k''c^2/2$, and we assume that the potential $V(F_0)$ is much slower than the abrupt jump in the $F_0$ state, so that we could neglect its influence, to finally get
 
\begin{align} 
 i\frac{\partial }{\partial T} \delta F=-D \frac{\partial^2}{\partial \eta^2}\delta F-i \Gamma \delta F-i\Gamma e^{2i \phi(\eta)}\delta F^*. 
\end{align} 

\section{Identifying the Fluctuation Equation as the Equation of Motion of $H$ in the Limit $\phi(\eta)=-q_0$}
We start with Eq.~\ref{eq:FluctuationEOM} and expand the field into bosonic annihilation and creation operators $\delta F=(1/\sqrt{2\pi})\int dq\, e^{iq\eta} \hat{a}_q$ and $\delta F^*=(1/\sqrt{2\pi})\int dq\, e^{-iq\eta} \hat{a}_q^\dagger$:
\begin{align}
    \begin{aligned}
    i\frac{\partial }{\partial T} \delta F&=-D \frac{\partial^2}{\partial \eta^2}\delta F-i \Gamma \delta F-i\Gamma e^{-2i q_0}\delta F^*\\
    i\frac{\partial}{\partial T}\int dq\, e^{iq\eta}\hat{a}_q &= (-D\frac{\partial^2}{\partial \eta^2}-i\Gamma)\int dq\, e^{iq\eta} \hat{a}_q-i\Gamma e^{-2iq_0}\int dq\, e^{-iq\eta} \hat{a}_q^\dagger
    \end{aligned}
\end{align}
Separating the Fourier modes, we get equations of motion for the bosonic annihilation and creation operators reading
\begin{align}
    \begin{aligned}
    i\frac{\partial}{\partial T}\hat{a}_q&=(Dq^2-i\Gamma)\hat{a}_q-i\Gamma \hat{a}_{-q-2q_0}^\dagger\, ,\\
    i\frac{\partial}{\partial T}\hat{a}_{-q-2q_0}^\dagger&=(-D(-q-2q_0)^2-i\Gamma)\hat{a}_{-q-2q_0}^\dagger-i\Gamma \hat{a}_{q}\,.
    \end{aligned}
\end{align}
Now shifting $q'=q+q_0$ and recasting these equations of motions into matrix form, we arrive at
\begin{align}
    i\frac{\partial}{\partial T} \begin{pmatrix}
        \hat{a}_{q'-q_0}\\
        \hat{a}_{-q'-q_0}^\dagger
    \end{pmatrix}
    =
    \begin{pmatrix}
        D(q-q_0)^2-i\Gamma & -i \Gamma\\
        -i \Gamma & D(-q-q_0)^2-i\Gamma
    \end{pmatrix}
    \begin{pmatrix}
        \hat{a}_{q'-q_0}\\
        \hat{a}_{-q'-q_0}^\dagger
    \end{pmatrix}\, .
    \label{eq:EOMDynamicalMatrix}
\end{align}
Identifying $D=1/4m$, we see that Eq.~\eqref{eq:EOMDynamicalMatrix}
corresponds to the equation of motion $i\dot{\mathbf{a}}_{q'}=\mathbf{D}_{q'}\mathbf{a}_{q'}$ \cite{Xiao2009TransformationTheory} generated by the dynamical matrix in Eq.~\eqref{eq:DynamicalMatrix}. The fluctuations on top of the extendon state in the limit $\phi(\eta)=-q_0$ are therefore described by the theoretical model in Eq.~\eqref{eq:MainHamiltonian}.

\section{Numerical Simulation}\label{secA3}

We base our numerical simulations of a modulated Fabry-P\'erot QCL on a Maxwell-Bloch derived mean-field theory~\cite{Burghoff2020d} of the form
\begin{align}
    \begin{aligned}
        \frac{\partial F}{\partial T} &= -\frac{1}{3}r\left(|F|^2+2\langle K \rangle^{-1}\Tilde{K}[|F|^2]-3P_0\right)F+i\frac{1}{2}\beta\frac{\partial^2F}{\partial z^2}+\frac{1}{4}D_g\frac{\partial^2F}{\partial z^2}\\
        &-\Tilde{K}\left[\gamma_1\frac{\partial F^*}{\partial z}F+\gamma_2F^*\frac{\partial F}{\partial z}\right]F - iM\cos{\left(Kz-\Delta\Omega T\right)}F
    \end{aligned}
\end{align}
with the slowly varying field $E(z,t)$ normalized using the effective gain to the quantity $F(z,t)$, which is continuous throughout the entire cavity round-trip as
\begin{align}
    F(z,t)= E(z,t)\exp{\left[-\frac{1}{2}\int_0^zg_{eff}(z')dz'\right]}= E(z,t)K^{-1/2}(z),
\end{align}
where the effective gain is defined as
\begin{align}
    g_{eff}(z)= g_0 -\alpha_w - g_0\frac{P(z)+2P(-z)}{P_s}+\ln{R}(\delta(z)+\delta(z-L_c)),
\end{align}
with the mirror losses $\alpha_w$, the steady-state power $P(z)=P_0K^{1/2}(z)$, with $P_0=P(0)$ the power at the facet and the mirror reflectivity $R$, which can be calculated using the refractive index contrast at the facet.The saturation power is given by
\begin{align}
    P_s= \frac{\hbar^2}{4|z_0e|^2T_1T_2},
\end{align}
and the small signal gain by
\begin{align}
    g_0=\frac{ez_0^2\omega_0JT_1T_2}{\epsilon_0cn\hbar L_{mod}}.
\end{align}
The dispersion is normalized as $\beta = k''\left(\frac{c}{n}\right)^3$ and the gain curvature is $D_g=2g_0T^2_2\left(\frac{c}{n}\right)^3$. The cross-steepening is characterized by the parameters $\gamma_1=\frac{g_0}{2P_s}\left(\frac{c}{n}\right)^2(2T_1+3T_2)$ and $\gamma_2=\frac{g_0}{2P_s}\left(\frac{c}{n}\right)^2(T_1+\frac{5}{2}T_2)$ and the energy relaxation by $r=3\frac{g_0}{2P_s}\frac{c}{n}\langle K \rangle$ with $\langle K \rangle=\frac{1}{2L_c}\int_0^{2L_c}K(u)du$. $\Tilde{K}[f](z)=\frac{1}{4L_c}\int_0^{4L_c}K(-\frac{u}{2})f(z-u)du$ is the convolution of the field with the effective gain of the counter-propagating field.

The external modulation is added as a phase modulation with modulation depth $M = \frac{g_0}{2}\frac{c}{n}\alpha\Delta J$ with the linewidth-enhancement factor $\alpha$ and the current modulation $\Delta J$.

\begin{table}[h]
\caption{Simulation parameters for all simulations used in the paper, when not explicitly used as a sweep variable.}\label{tab1}
\begin{tabular}{@{}lll@{}}
\toprule
Name & Symbol & Value\\
\midrule
Cavity length    & $L_c$    & $4 \text{ mm}$    \\
Refractive index    & $n$    & 3.3 \\
Mirror reflectivity    & $R$    & 0.286 \\
Waveguide losses    & $\alpha_w$    & $4\text{ cm}^{-1}$ \\
Dipole moment    & $z_0$    & $2.3\text{ nm}$ \\
Population Lifetime    & $T_1$    & $0.4\text{ ps}$ \\
Coherence Lifetime    & $T_2$    & $50\text{ fs}$ \\
GVD    & $\frac{k''}{2L_c}$    & $-1500 \text{ fs}^2 \text{ mm}^{-1}$ \\
Linewidth enhancement factor    & $\alpha$    & 0.5 \\
Current density     & $J$    & $1.725\text{ kA cm}^{-2}$ \\
Module length & $L_{mod}$ & $58\text{ nm}$\\
Node spacing & $\Delta z$ & $8\text{ \textmu m}$\\
Kerr nonlinearity & $\gamma_k$ & $-35\text{ pm V}^{-2}$ \\
Current density modulation & $\Delta J$ & $50\text{ A cm}^{-2}$ \\
\botrule
\end{tabular}
\end{table}
\newpage

\section{Optical Spectra used for Calculating the Transform-Limit}\label{secA4}
The Fourier-limit for the experimental data was calculated as the inverse Fourier-transform using the spectrum obtained from a Fourier-transform spectrometer. Since in a standard interferogram all phase information coming from the field is lost, the result from an inverse Fourier Transform corresponds to the signal with perfectly flat phases.
\begin{figure*}[htbp]
\centering
\includegraphics[width=\textwidth]{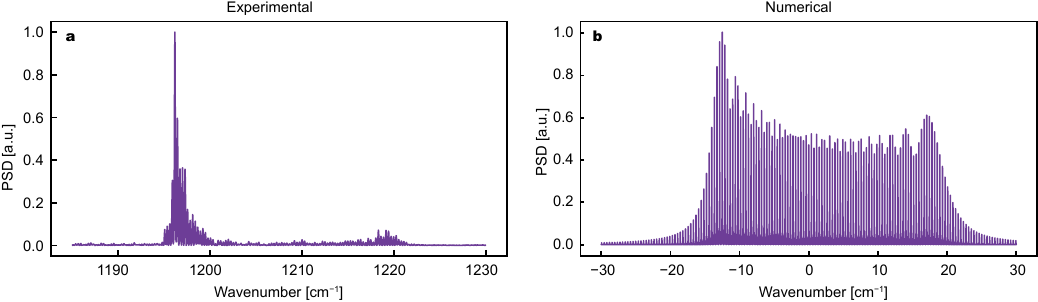}
\caption{Optical spectra used for calculating the Transform-limited pulses. \textbf{a}, Experimentally measured spectrum using a Fourier-Transform Spectrometer at the same parameters as used during the ASUPS measuement. \textbf{b}, Spectrum generated for the result of the numerical simulations.}
\label{figA1}
\end{figure*} 

\end{appendices}

%% BioMed_Central_Bib_Style_v1.01


\begin{thebibliography}{92}
% BibTex style file: bmc-mathphys.bst (version 2.1), 2014-07-24
\ifx \bisbn   \undefined \def \bisbn  #1{ISBN #1}\fi
\ifx \binits  \undefined \def \binits#1{#1}\fi
\ifx \bauthor  \undefined \def \bauthor#1{#1}\fi
\ifx \batitle  \undefined \def \batitle#1{#1}\fi
\ifx \bjtitle  \undefined \def \bjtitle#1{#1}\fi
\ifx \bvolume  \undefined \def \bvolume#1{\textbf{#1}}\fi
\ifx \byear  \undefined \def \byear#1{#1}\fi
\ifx \bissue  \undefined \def \bissue#1{#1}\fi
\ifx \bfpage  \undefined \def \bfpage#1{#1}\fi
\ifx \blpage  \undefined \def \blpage #1{#1}\fi
\ifx \burl  \undefined \def \burl#1{\textsf{#1}}\fi
\ifx \doiurl  \undefined \def \doiurl#1{\url{https://doi.org/#1}}\fi
\ifx \betal  \undefined \def \betal{\textit{et al.}}\fi
\ifx \binstitute  \undefined \def \binstitute#1{#1}\fi
\ifx \binstitutionaled  \undefined \def \binstitutionaled#1{#1}\fi
\ifx \bctitle  \undefined \def \bctitle#1{#1}\fi
\ifx \beditor  \undefined \def \beditor#1{#1}\fi
\ifx \bpublisher  \undefined \def \bpublisher#1{#1}\fi
\ifx \bbtitle  \undefined \def \bbtitle#1{#1}\fi
\ifx \bedition  \undefined \def \bedition#1{#1}\fi
\ifx \bseriesno  \undefined \def \bseriesno#1{#1}\fi
\ifx \blocation  \undefined \def \blocation#1{#1}\fi
\ifx \bsertitle  \undefined \def \bsertitle#1{#1}\fi
\ifx \bsnm \undefined \def \bsnm#1{#1}\fi
\ifx \bsuffix \undefined \def \bsuffix#1{#1}\fi
\ifx \bparticle \undefined \def \bparticle#1{#1}\fi
\ifx \barticle \undefined \def \barticle#1{#1}\fi
\bibcommenthead
\ifx \bconfdate \undefined \def \bconfdate #1{#1}\fi
\ifx \botherref \undefined \def \botherref #1{#1}\fi
\ifx \url \undefined \def \url#1{\textsf{#1}}\fi
\ifx \bchapter \undefined \def \bchapter#1{#1}\fi
\ifx \bbook \undefined \def \bbook#1{#1}\fi
\ifx \bcomment \undefined \def \bcomment#1{#1}\fi
\ifx \oauthor \undefined \def \oauthor#1{#1}\fi
\ifx \citeauthoryear \undefined \def \citeauthoryear#1{#1}\fi
\ifx \endbibitem  \undefined \def \endbibitem {}\fi
\ifx \bconflocation  \undefined \def \bconflocation#1{#1}\fi
\ifx \arxivurl  \undefined \def \arxivurl#1{\textsf{#1}}\fi
\csname PreBibitemsHook\endcsname

%%% 1
\bibitem[\protect\citeauthoryear{Nakahara}{2018}]{Nakahara2018GeometryPhysics}
\begin{bbook}
\bauthor{\bsnm{Nakahara}, \binits{M.}}:
\bbtitle{{Geometry, Topology and Physics}},
\bedition{2}nd edn.
\bpublisher{CRC Press},
\blocation{Boca Raton}
(\byear{2018}).
\doiurl{10.1201/9781315275826} .
\burl{https://www.taylorfrancis.com/books/9781420056945}
\end{bbook}
\endbibitem

%%% 2
\bibitem[\protect\citeauthoryear{Bernevig}{2013}]{Bernevig2013TopologicalSuperconductors}
\begin{bbook}
\bauthor{\bsnm{Bernevig}, \binits{B.A.}}:
\bbtitle{{Topological Insulators and Topological Superconductors}},
p. \bfpage{247}.
\bpublisher{Princeton University Press},
\blocation{Princeton}
(\byear{2013}).
\doiurl{10.1515/9781400846733} .
\burl{https://www.degruyter.com/document/doi/10.1515/9781400846733/html}
\end{bbook}
\endbibitem

%%% 3
\bibitem[\protect\citeauthoryear{Hasan and Kane}{2010}]{Hasan2010ColloquiumInsulators}
\begin{barticle}
\bauthor{\bsnm{Hasan}, \binits{M.Z.}},
\bauthor{\bsnm{Kane}, \binits{C.L.}}:
\batitle{{Colloquium : Topological insulators}}.
\bjtitle{Reviews of Modern Physics}
\bvolume{82}(\bissue{4}),
\bfpage{3045}--\blpage{3067}
(\byear{2010})
\doiurl{10.1103/RevModPhys.82.3045}
\end{barticle}
\endbibitem

%%% 4
\bibitem[\protect\citeauthoryear{Qi and Zhang}{2011}]{qi2011topological}
\begin{barticle}
\bauthor{\bsnm{Qi}, \binits{X.-L.}},
\bauthor{\bsnm{Zhang}, \binits{S.-C.}}:
\batitle{Topological insulators and superconductors}.
\bjtitle{Reviews of modern physics}
\bvolume{83}(\bissue{4}),
\bfpage{1057}--\blpage{1110}
(\byear{2011})
\end{barticle}
\endbibitem

%%% 5
\bibitem[\protect\citeauthoryear{Prange et~al.}{2012}]{QuantumHallEffectBook}
\begin{bbook}
\bauthor{\bsnm{Prange}, \binits{R.E.}},
\bauthor{\bsnm{Cage}, \binits{M.E.}},
\bauthor{\bsnm{Thouless}, \binits{D.J.}},
\bauthor{\bsnm{Pruisken}, \binits{A.M.M.}},
\bauthor{\bsnm{Chang}, \binits{A.M.}},
\bauthor{\bsnm{Laughlin}, \binits{R.B.}},
\bauthor{\bsnm{Duncan}, \binits{F.}},
\bauthor{\bsnm{Haldane}, \binits{M.}},
\bauthor{\bsnm{Girvin}, \binits{S.M.}}:
\bbtitle{The Quantum Hall Effect},
\bedition{2}nd edn.
\bpublisher{Springer},
\blocation{New York}
(\byear{2012}).
\doiurl{10.1007/978-1-4612-3350-3} .
\burl{https://link.springer.com/book/10.1007/978-1-4612-3350-3}
\end{bbook}
\endbibitem

%%% 6
\bibitem[\protect\citeauthoryear{Haldane}{1988}]{Haldane1988ParityAnomaly}
\begin{barticle}
\bauthor{\bsnm{Haldane}, \binits{F.D.M.}}:
\batitle{Model for a quantum hall effect without landau levels: Condensed-matter realization of the "parity anomaly"}.
\bjtitle{Phys. Rev. Lett.}
\bvolume{61},
\bfpage{2015}--\blpage{2018}
(\byear{1988})
\doiurl{10.1103/PhysRevLett.61.2015}
\end{barticle}
\endbibitem

%%% 7
\bibitem[\protect\citeauthoryear{Hsieh et~al.}{2008}]{Hsieh2008DiracInsulator}
\begin{barticle}
\bauthor{\bsnm{Hsieh}, \binits{D.}},
\bauthor{\bsnm{Qian}, \binits{D.}},
\bauthor{\bsnm{Wray}, \binits{L.}},
\bauthor{\bsnm{Xia}, \binits{Y.}},
\bauthor{\bsnm{Hor}, \binits{Y.S.}},
\bauthor{\bsnm{.Cava}, \binits{R.J.}},
\bauthor{\bsnm{Hasan}, \binits{M.Z.}}:
\batitle{A topological dirac insulator in a quantum spin hall phase}.
\bjtitle{Nature}
\bvolume{452},
\bfpage{970}--\blpage{974}
(\byear{2008})
\doiurl{10.1038/nature06843}
\end{barticle}
\endbibitem

%%% 8
\bibitem[\protect\citeauthoryear{Fu and Kane}{2008}]{Liang2008SuperconductingProximity}
\begin{barticle}
\bauthor{\bsnm{Fu}, \binits{L.}},
\bauthor{\bsnm{Kane}, \binits{C.L.}}:
\batitle{Superconducting proximity effect and majorana fermions at the surface of a topological insulator}.
\bjtitle{Phys. Rev. Lett.}
\bvolume{100},
\bfpage{096407}
(\byear{2008})
\doiurl{10.1103/PhysRevLett.100.096407}
\end{barticle}
\endbibitem

%%% 9
\bibitem[\protect\citeauthoryear{Xu et~al.}{2015}]{Xu2015FermiArcs}
\begin{barticle}
\bauthor{\bsnm{Xu}, \binits{S.-Y.}},
\bauthor{\bsnm{Liu}, \binits{C.}},
\bauthor{\bsnm{Kushwaha}, \binits{S.K.}},
\bauthor{\bsnm{Sankar}, \binits{R.}},
\bauthor{\bsnm{Krizan}, \binits{J.W.}},
\bauthor{\bsnm{Belopolski}, \binits{I.}},
\bauthor{\bsnm{Neupane}, \binits{M.}},
\bauthor{\bsnm{Bian}, \binits{G.}},
\bauthor{\bsnm{Alidoust}, \binits{N.}},
\bauthor{\bsnm{Chang}, \binits{T.-R.}},
\bauthor{\bsnm{Jeng}, \binits{H.-T.}},
\bauthor{\bsnm{Huang}, \binits{C.-Y.}},
\bauthor{\bsnm{Tsai}, \binits{W.-F.}},
\bauthor{\bsnm{Lin}, \binits{H.}},
\bauthor{\bsnm{Shibayev}, \binits{P.P.}},
\bauthor{\bsnm{Chou}, \binits{F.-C.}},
\bauthor{\bsnm{Cava}, \binits{R.J.}},
\bauthor{\bsnm{Hasan}, \binits{M.Z.}}:
\batitle{Observation of fermi arc surface states in a topological metal}.
\bjtitle{Science}
\bvolume{347}(\bissue{6219}),
\bfpage{294}--\blpage{298}
(\byear{2015})
\doiurl{10.1126/science.1256742}
{\href{https://arxiv.org/abs/https://www.science.org/doi/pdf/10.1126/science.1256742}{{https://www.science.org/doi/pdf/10.1126/science.1256742}}}
\end{barticle}
\endbibitem

%%% 10
\bibitem[\protect\citeauthoryear{Wang et~al.}{2008}]{Wang2008OneWayEdgeModes}
\begin{barticle}
\bauthor{\bsnm{Wang}, \binits{Z.}},
\bauthor{\bsnm{Chong}, \binits{Y.D.}},
\bauthor{\bsnm{Joannopoulos}, \binits{J.D.}},
\bauthor{\bsnm{Solja\v{c}i\'{c}}, \binits{M.}}:
\batitle{Reflection-free one-way edge modes in a gyromagnetic photonic crystal}.
\bjtitle{Phys. Rev. Lett.}
\bvolume{100},
\bfpage{013905}
(\byear{2008})
\doiurl{10.1103/PhysRevLett.100.013905}
\end{barticle}
\endbibitem

%%% 11
\bibitem[\protect\citeauthoryear{Khanikaev et~al.}{2013}]{Khanikaev2013PhotonicTopoInsu}
\begin{barticle}
\bauthor{\bsnm{Khanikaev}, \binits{A.B.}},
\bauthor{\bsnm{Hossein~Mousavi}, \binits{S.}},
\bauthor{\bsnm{Tse}, \binits{W.-K.}},
\bauthor{\bsnm{Kargarian}, \binits{M.}},
\bauthor{\bsnm{MacDonald}, \binits{A.H.}},
\bauthor{\bsnm{Shvets}, \binits{G.}}:
\batitle{Photonic topological insulators}.
\bjtitle{Nature Materials}
\bvolume{12}(\bissue{3}),
\bfpage{233}--\blpage{239}
(\byear{2013})
\doiurl{10.1038/nmat3520}
\end{barticle}
\endbibitem

%%% 12
\bibitem[\protect\citeauthoryear{Rechtsman et~al.}{2013}]{Rechtsman2013FloquetTopoInsu}
\begin{barticle}
\bauthor{\bsnm{Rechtsman}, \binits{M.C.}},
\bauthor{\bsnm{Zeuner}, \binits{J.M.}},
\bauthor{\bsnm{Plotnik}, \binits{Y.}},
\bauthor{\bsnm{Lumer}, \binits{Y.}},
\bauthor{\bsnm{Podolsky}, \binits{D.}},
\bauthor{\bsnm{Dreisow}, \binits{F.}},
\bauthor{\bsnm{Nolte}, \binits{S.}},
\bauthor{\bsnm{Segev}, \binits{M.}},
\bauthor{\bsnm{Szameit}, \binits{A.}}:
\batitle{Photonic floquet topological insulators}.
\bjtitle{Nature}
\bvolume{496}(\bissue{7444}),
\bfpage{196}--\blpage{200}
(\byear{2013})
\doiurl{10.1038/nature12066}
\end{barticle}
\endbibitem

%%% 13
\bibitem[\protect\citeauthoryear{Hafezi et~al.}{2013}]{Hafezi2013ImagingEdgeState}
\begin{barticle}
\bauthor{\bsnm{Hafezi}, \binits{M.}},
\bauthor{\bsnm{Mittal}, \binits{S.}},
\bauthor{\bsnm{Fan}, \binits{J.}},
\bauthor{\bsnm{Migdall}, \binits{A.}},
\bauthor{\bsnm{Taylor}, \binits{J.M.}}:
\batitle{Imaging topological edge states in silicon photonics}.
\bjtitle{Nature Photonics}
\bvolume{7}(\bissue{12}),
\bfpage{1001}--\blpage{1005}
(\byear{2013})
\doiurl{10.1038/nphoton.2013.274}
\end{barticle}
\endbibitem

%%% 14
\bibitem[\protect\citeauthoryear{Süsstrunk and Huber}{2015}]{Susstrunk2015MechanicalTopoInsu}
\begin{barticle}
\bauthor{\bsnm{Süsstrunk}, \binits{R.}},
\bauthor{\bsnm{Huber}, \binits{S.D.}}:
\batitle{Observation of phononic helical edge states in a mechanical topological insulator}.
\bjtitle{Science}
\bvolume{349}(\bissue{6243}),
\bfpage{47}--\blpage{50}
(\byear{2015})
\doiurl{10.1126/science.aab0239}
{\href{https://arxiv.org/abs/https://www.science.org/doi/pdf/10.1126/science.aab0239}{{https://www.science.org/doi/pdf/10.1126/science.aab0239}}}
\end{barticle}
\endbibitem

%%% 15
\bibitem[\protect\citeauthoryear{Ozawa et~al.}{2019}]{Ozawa2019TopologicalPhotonics}
\begin{barticle}
\bauthor{\bsnm{Ozawa}, \binits{T.}},
\bauthor{\bsnm{Price}, \binits{H.M.}},
\bauthor{\bsnm{Amo}, \binits{A.}},
\bauthor{\bsnm{Goldman}, \binits{N.}},
\bauthor{\bsnm{Hafezi}, \binits{M.}},
\bauthor{\bsnm{Lu}, \binits{L.}},
\bauthor{\bsnm{Rechtsman}, \binits{M.C.}},
\bauthor{\bsnm{Schuster}, \binits{D.}},
\bauthor{\bsnm{Simon}, \binits{J.}},
\bauthor{\bsnm{Zilberberg}, \binits{O.}},
\bauthor{\bsnm{Carusotto}, \binits{I.}}:
\batitle{{Topological photonics}}.
\bjtitle{Reviews of Modern Physics}
\bvolume{91}(\bissue{1}),
\bfpage{015006}
(\byear{2019})
\doiurl{10.1103/RevModPhys.91.015006}
\end{barticle}
\endbibitem

%%% 16
\bibitem[\protect\citeauthoryear{Price et~al.}{2022}]{Price2022RoadmapPhotonics}
\begin{barticle}
\bauthor{\bsnm{Price}, \binits{H.}},
\bauthor{\bsnm{Chong}, \binits{Y.}},
\bauthor{\bsnm{Khanikaev}, \binits{A.}},
\bauthor{\bsnm{Schomerus}, \binits{H.}},
\bauthor{\bsnm{Maczewsky}, \binits{L.J.}},
\bauthor{\bsnm{Kremer}, \binits{M.}},
\bauthor{\bsnm{Heinrich}, \binits{M.}},
\bauthor{\bsnm{Szameit}, \binits{A.}},
\bauthor{\bsnm{Zilberberg}, \binits{O.}},
\bauthor{\bsnm{Yang}, \binits{Y.}},
\bauthor{\bsnm{Zhang}, \binits{B.}},
\bauthor{\bsnm{Al{\`{u}}}, \binits{A.}},
\bauthor{\bsnm{Thomale}, \binits{R.}},
\bauthor{\bsnm{Carusotto}, \binits{I.}},
\bauthor{\bsnm{St-Jean}, \binits{P.}},
\bauthor{\bsnm{Amo}, \binits{A.}},
\bauthor{\bsnm{Dutt}, \binits{A.}},
\bauthor{\bsnm{Yuan}, \binits{L.}},
\bauthor{\bsnm{Fan}, \binits{S.}},
\bauthor{\bsnm{Yin}, \binits{X.}},
\bauthor{\bsnm{Peng}, \binits{C.}},
\bauthor{\bsnm{Ozawa}, \binits{T.}},
\bauthor{\bsnm{Blanco-Redondo}, \binits{A.}}:
\batitle{{Roadmap on topological photonics}}.
\bjtitle{Journal of Physics: Photonics}
\bvolume{4}(\bissue{3}),
\bfpage{032501}
(\byear{2022})
\doiurl{10.1088/2515-7647/ac4ee4}
\end{barticle}
\endbibitem

%%% 17
\bibitem[\protect\citeauthoryear{Lu et~al.}{2014}]{Lu2014TopoPhotonics}
\begin{barticle}
\bauthor{\bsnm{Lu}, \binits{L.}},
\bauthor{\bsnm{Joannopoulos}, \binits{J.D.}},
\bauthor{\bsnm{Solja{\v{c}}i{\'{c}}}, \binits{M.}}:
\batitle{Topological photonics}.
\bjtitle{Nature Photonics}
\bvolume{8}(\bissue{11}),
\bfpage{821}--\blpage{829}
(\byear{2014})
\doiurl{10.1038/nphoton.2014.248}
\end{barticle}
\endbibitem

%%% 18
\bibitem[\protect\citeauthoryear{Kraus et~al.}{2012}]{Kraus2012QuasicrystalPumping}
\begin{barticle}
\bauthor{\bsnm{Kraus}, \binits{Y.E.}},
\bauthor{\bsnm{Lahini}, \binits{Y.}},
\bauthor{\bsnm{Ringel}, \binits{Z.}},
\bauthor{\bsnm{Verbin}, \binits{M.}},
\bauthor{\bsnm{Zilberberg}, \binits{O.}}:
\batitle{Topological states and adiabatic pumping in quasicrystals}.
\bjtitle{Phys. Rev. Lett.}
\bvolume{109},
\bfpage{106402}
(\byear{2012})
\doiurl{10.1103/PhysRevLett.109.106402}
\end{barticle}
\endbibitem

%%% 19
\bibitem[\protect\citeauthoryear{Zilberberg et~al.}{2018}]{Zilberberg20184DQuantumHallPumping}
\begin{barticle}
\bauthor{\bsnm{Zilberberg}, \binits{O.}},
\bauthor{\bsnm{Huang}, \binits{S.}},
\bauthor{\bsnm{Guglielmon}, \binits{J.}},
\bauthor{\bsnm{Wang}, \binits{M.}},
\bauthor{\bsnm{Chen}, \binits{K.P.}},
\bauthor{\bsnm{Kraus}, \binits{Y.E.}},
\bauthor{\bsnm{Rechtsman}, \binits{M.C.}}:
\batitle{Photonic topological boundary pumping as a probe of 4d quantum hall physics}.
\bjtitle{Nature}
\bvolume{553}(\bissue{7686}),
\bfpage{59}--\blpage{62}
(\byear{2018})
\doiurl{10.1038/nature25011}
\end{barticle}
\endbibitem

%%% 20
\bibitem[\protect\citeauthoryear{Kremer et~al.}{2020}]{Kremer202SquareRoot}
\begin{barticle}
\bauthor{\bsnm{Kremer}, \binits{M.}},
\bauthor{\bsnm{Petrides}, \binits{I.}},
\bauthor{\bsnm{Meyer}, \binits{E.}},
\bauthor{\bsnm{Heinrich}, \binits{M.}},
\bauthor{\bsnm{Zilberberg}, \binits{O.}},
\bauthor{\bsnm{Szameit}, \binits{A.}}:
\batitle{A square-root topological insulator with non-quantized indices realized with photonic aharonov-bohm cages}.
\bjtitle{Nature Communications}
\bvolume{11}(\bissue{1}),
\bfpage{907}
(\byear{2020})
\doiurl{10.1038/s41467-020-14692-4}
\end{barticle}
\endbibitem

%%% 21
\bibitem[\protect\citeauthoryear{Noh et~al.}{2018}]{Noh2018MidGapModes}
\begin{barticle}
\bauthor{\bsnm{Noh}, \binits{J.}},
\bauthor{\bsnm{Benalcazar}, \binits{W.A.}},
\bauthor{\bsnm{Huang}, \binits{S.}},
\bauthor{\bsnm{Collins}, \binits{M.J.}},
\bauthor{\bsnm{Chen}, \binits{K.P.}},
\bauthor{\bsnm{Hughes}, \binits{T.L.}},
\bauthor{\bsnm{Rechtsman}, \binits{M.C.}}:
\batitle{Topological protection of photonic mid-gap defect modes}.
\bjtitle{Nature Photonics}
\bvolume{12}(\bissue{7}),
\bfpage{408}--\blpage{415}
(\byear{2018})
\doiurl{10.1038/s41566-018-0179-3}
\end{barticle}
\endbibitem

%%% 22
\bibitem[\protect\citeauthoryear{Tambasco et~al.}{2018}]{tambasco2018quantum}
\begin{barticle}
\bauthor{\bsnm{Tambasco}, \binits{J.-L.}},
\bauthor{\bsnm{Corrielli}, \binits{G.}},
\bauthor{\bsnm{Chapman}, \binits{R.J.}},
\bauthor{\bsnm{Crespi}, \binits{A.}},
\bauthor{\bsnm{Zilberberg}, \binits{O.}},
\bauthor{\bsnm{Osellame}, \binits{R.}},
\bauthor{\bsnm{Peruzzo}, \binits{A.}}:
\batitle{Quantum interference of topological states of light}.
\bjtitle{Science advances}
\bvolume{4}(\bissue{9}),
\bfpage{3187}
(\byear{2018})
\end{barticle}
\endbibitem

%%% 23
\bibitem[\protect\citeauthoryear{Yan et~al.}{2023}]{Yan2023AdvancesPhotonics}
\begin{barticle}
\bauthor{\bsnm{Yan}, \binits{Q.}},
\bauthor{\bsnm{Zhao}, \binits{B.}},
\bauthor{\bsnm{Zhou}, \binits{R.}},
\bauthor{\bsnm{Ma}, \binits{R.}},
\bauthor{\bsnm{Lyu}, \binits{Q.}},
\bauthor{\bsnm{Chu}, \binits{S.}},
\bauthor{\bsnm{Hu}, \binits{X.}},
\bauthor{\bsnm{Gong}, \binits{Q.}}:
\batitle{Advances and applications on non-hermitian topological photonics}.
\bjtitle{Nanophotonics}
\bvolume{12}(\bissue{13}),
\bfpage{2247}--\blpage{2271}
(\byear{2023})
\doiurl{10.1515/nanoph-2022-0775}
\end{barticle}
\endbibitem

%%% 24
\bibitem[\protect\citeauthoryear{Lieu}{2018}]{Lieu2018TopologicalModel}
\begin{barticle}
\bauthor{\bsnm{Lieu}, \binits{S.}}:
\batitle{{Topological phases in the non-Hermitian Su-Schrieffer-Heeger model}}.
\bjtitle{Physical Review B}
\bvolume{97}(\bissue{4}),
\bfpage{045106}
(\byear{2018})
\doiurl{10.1103/PhysRevB.97.045106}
\end{barticle}
\endbibitem

%%% 25
\bibitem[\protect\citeauthoryear{Zeuner et~al.}{2015}]{Zeuner2015ObservationSystem}
\begin{barticle}
\bauthor{\bsnm{Zeuner}, \binits{J.M.}},
\bauthor{\bsnm{Rechtsman}, \binits{M.C.}},
\bauthor{\bsnm{Plotnik}, \binits{Y.}},
\bauthor{\bsnm{Lumer}, \binits{Y.}},
\bauthor{\bsnm{Nolte}, \binits{S.}},
\bauthor{\bsnm{Rudner}, \binits{M.S.}},
\bauthor{\bsnm{Segev}, \binits{M.}},
\bauthor{\bsnm{Szameit}, \binits{A.}}:
\batitle{{Observation of a Topological Transition in the Bulk of a Non-Hermitian System}}.
\bjtitle{Physical Review Letters}
\bvolume{115}(\bissue{4}),
\bfpage{040402}
(\byear{2015})
\doiurl{10.1103/PhysRevLett.115.040402}
\end{barticle}
\endbibitem

%%% 26
\bibitem[\protect\citeauthoryear{Hu and Hughes}{2011}]{Hu2011AbsenceHamiltonians}
\begin{barticle}
\bauthor{\bsnm{Hu}, \binits{Y.C.}},
\bauthor{\bsnm{Hughes}, \binits{T.L.}}:
\batitle{{Absence of topological insulator phases in non-Hermitian <math display="inline"> <mrow> <mi>P</mi> <mi>T</mi> </mrow> </math> -symmetric Hamiltonians}}.
\bjtitle{Physical Review B}
\bvolume{84}(\bissue{15}),
\bfpage{153101}
(\byear{2011})
\doiurl{10.1103/PhysRevB.84.153101}
\end{barticle}
\endbibitem

%%% 27
\bibitem[\protect\citeauthoryear{Weimann et~al.}{2017}]{Weimann2017TopologicallyCrystals}
\begin{barticle}
\bauthor{\bsnm{Weimann}, \binits{S.}},
\bauthor{\bsnm{Kremer}, \binits{M.}},
\bauthor{\bsnm{Plotnik}, \binits{Y.}},
\bauthor{\bsnm{Lumer}, \binits{Y.}},
\bauthor{\bsnm{Nolte}, \binits{S.}},
\bauthor{\bsnm{Makris}, \binits{K.G.}},
\bauthor{\bsnm{Segev}, \binits{M.}},
\bauthor{\bsnm{Rechtsman}, \binits{M.C.}},
\bauthor{\bsnm{Szameit}, \binits{A.}}:
\batitle{{Topologically protected bound states in photonic parity–time-symmetric crystals}}.
\bjtitle{Nature Materials}
\bvolume{16}(\bissue{4}),
\bfpage{433}--\blpage{438}
(\byear{2017})
\doiurl{10.1038/nmat4811}
\end{barticle}
\endbibitem

%%% 28
\bibitem[\protect\citeauthoryear{Yao and Wang}{2018}]{Yao2018EdgeStatesInNHSystems}
\begin{barticle}
\bauthor{\bsnm{Yao}, \binits{S.}},
\bauthor{\bsnm{Wang}, \binits{Z.}}:
\batitle{Edge states and topological invariants of non-hermitian systems}.
\bjtitle{Phys. Rev. Lett.}
\bvolume{121},
\bfpage{086803}
(\byear{2018})
\doiurl{10.1103/PhysRevLett.121.086803}
\end{barticle}
\endbibitem

%%% 29
\bibitem[\protect\citeauthoryear{Su et~al.}{2021}]{Su2021DirectSystem}
\begin{barticle}
\bauthor{\bsnm{Su}, \binits{R.}},
\bauthor{\bsnm{Estrecho}, \binits{E.}},
\bauthor{\bsnm{Biega{\'{n}}ska}, \binits{D.}},
\bauthor{\bsnm{Huang}, \binits{Y.}},
\bauthor{\bsnm{Wurdack}, \binits{M.}},
\bauthor{\bsnm{Pieczarka}, \binits{M.}},
\bauthor{\bsnm{Truscott}, \binits{A.G.}},
\bauthor{\bsnm{Liew}, \binits{T.C.H.}},
\bauthor{\bsnm{Ostrovskaya}, \binits{E.A.}},
\bauthor{\bsnm{Xiong}, \binits{Q.}}:
\batitle{{Direct measurement of a non-Hermitian topological invariant in a hybrid light-matter system}}.
\bjtitle{Science Advances}
\bvolume{7}(\bissue{45}),
\bfpage{8905}
(\byear{2021})
\doiurl{10.1126/sciadv.abj8905}
\end{barticle}
\endbibitem

%%% 30
\bibitem[\protect\citeauthoryear{Zheng et~al.}{2024}]{Zheng2024DynamicDimensions}
\begin{barticle}
\bauthor{\bsnm{Zheng}, \binits{X.}},
\bauthor{\bsnm{Jalali~Mehrabad}, \binits{M.}},
\bauthor{\bsnm{Vannucci}, \binits{J.}},
\bauthor{\bsnm{Li}, \binits{K.}},
\bauthor{\bsnm{Dutt}, \binits{A.}},
\bauthor{\bsnm{Hafezi}, \binits{M.}},
\bauthor{\bsnm{Mittal}, \binits{S.}},
\bauthor{\bsnm{Waks}, \binits{E.}}:
\batitle{{Dynamic control of 2D non-Hermitian photonic corner skin modes in synthetic dimensions}}.
\bjtitle{Nature Communications 2024 15:1}
\bvolume{15}(\bissue{1}),
\bfpage{1}--\blpage{7}
(\byear{2024})
\doiurl{10.1038/s41467-024-55236-4}
\end{barticle}
\endbibitem

%%% 31
\bibitem[\protect\citeauthoryear{Li et~al.}{2024}]{Li2024ObservationEffects}
\begin{barticle}
\bauthor{\bsnm{Li}, \binits{Z.}},
\bauthor{\bsnm{Wang}, \binits{L.W.}},
\bauthor{\bsnm{Wang}, \binits{X.}},
\bauthor{\bsnm{Lin}, \binits{Z.K.}},
\bauthor{\bsnm{Ma}, \binits{G.}},
\bauthor{\bsnm{Jiang}, \binits{J.H.}}:
\batitle{{Observation of dynamic non-Hermitian skin effects}}.
\bjtitle{Nature Communications 2024 15:1}
\bvolume{15}(\bissue{1}),
\bfpage{1}--\blpage{9}
(\byear{2024})
\doiurl{10.1038/s41467-024-50776-1}
\end{barticle}
\endbibitem

%%% 32
\bibitem[\protect\citeauthoryear{Fritzsche et~al.}{2024}]{Fritzsche2024Paritytime-symmetricInsulator}
\begin{barticle}
\bauthor{\bsnm{Fritzsche}, \binits{A.}},
\bauthor{\bsnm{Biesenthal}, \binits{T.}},
\bauthor{\bsnm{Maczewsky}, \binits{L.J.}},
\bauthor{\bsnm{Becker}, \binits{K.}},
\bauthor{\bsnm{Ehrhardt}, \binits{M.}},
\bauthor{\bsnm{Heinrich}, \binits{M.}},
\bauthor{\bsnm{Thomale}, \binits{R.}},
\bauthor{\bsnm{Joglekar}, \binits{Y.N.}},
\bauthor{\bsnm{Szameit}, \binits{A.}}:
\batitle{{Parity–time-symmetric photonic topological insulator}}.
\bjtitle{Nature Materials}
\bvolume{23}(\bissue{3}),
\bfpage{377}--\blpage{382}
(\byear{2024})
\doiurl{10.1038/s41563-023-01773-0}
\end{barticle}
\endbibitem

%%% 33
\bibitem[\protect\citeauthoryear{Zhao et~al.}{2025}]{Zhao2025Two-dimensionalGas}
\begin{barticle}
\bauthor{\bsnm{Zhao}, \binits{E.}},
\bauthor{\bsnm{Wang}, \binits{Z.}},
\bauthor{\bsnm{He}, \binits{C.}},
\bauthor{\bsnm{Poon}, \binits{T.F.J.}},
\bauthor{\bsnm{Pak}, \binits{K.K.}},
\bauthor{\bsnm{Liu}, \binits{Y.-J.}},
\bauthor{\bsnm{Ren}, \binits{P.}},
\bauthor{\bsnm{Liu}, \binits{X.-J.}},
\bauthor{\bsnm{Jo}, \binits{G.-B.}}:
\batitle{{Two-dimensional non-Hermitian skin effect in an ultracold Fermi gas}}.
\bjtitle{Nature 2025 637:8046}
\bvolume{637}(\bissue{8046}),
\bfpage{565}--\blpage{573}
(\byear{2025})
\doiurl{10.1038/s41586-024-08347-3}
\end{barticle}
\endbibitem

%%% 34
\bibitem[\protect\citeauthoryear{Okuma et~al.}{2020}]{Okuma2019TopologicalNHSE}
\begin{barticle}
\bauthor{\bsnm{Okuma}, \binits{N.}},
\bauthor{\bsnm{Kawabata}, \binits{K.}},
\bauthor{\bsnm{Shiozaki}, \binits{K.}},
\bauthor{\bsnm{Sato}, \binits{M.}}:
\batitle{Topological origin of non-hermitian skin effects}.
\bjtitle{Phys. Rev. Lett.}
\bvolume{124},
\bfpage{086801}
(\byear{2020})
\doiurl{10.1103/PhysRevLett.124.086801}
\end{barticle}
\endbibitem

%%% 35
\bibitem[\protect\citeauthoryear{Hatano and Nelson}{1996}]{Hatano1996LocalizationMechanics}
\begin{barticle}
\bauthor{\bsnm{Hatano}, \binits{N.}},
\bauthor{\bsnm{Nelson}, \binits{D.R.}}:
\batitle{{Localization Transitions in Non-Hermitian Quantum Mechanics}}.
\bjtitle{Physical Review Letters}
\bvolume{77}(\bissue{3}),
\bfpage{570}--\blpage{573}
(\byear{1996})
\doiurl{10.1103/PhysRevLett.77.570}
\end{barticle}
\endbibitem

%%% 36
\bibitem[\protect\citeauthoryear{Bestler et~al.}{2025}]{bestlerModel}
\begin{botherref}
\oauthor{\bsnm{Bestler}, \binits{M.}},
\oauthor{\bsnm{Dikopoltsev}, \binits{A.}},
\oauthor{\bsnm{Zilberberg}, \binits{O.}}:
Non-Hermitian topology and skin modes in the continuum via parametric processes
(2025).
\url{https://arxiv.org/abs/2505.02776}
\end{botherref}
\endbibitem

%%% 37
\bibitem[\protect\citeauthoryear{Wanjura et~al.}{2020}]{Wanjura2020TopologicalArrays}
\begin{barticle}
\bauthor{\bsnm{Wanjura}, \binits{C.C.}},
\bauthor{\bsnm{Brunelli}, \binits{M.}},
\bauthor{\bsnm{Nunnenkamp}, \binits{A.}}:
\batitle{{Topological framework for directional amplification in driven-dissipative cavity arrays}}.
\bjtitle{Nature Communications}
\bvolume{11}(\bissue{1}),
\bfpage{3149}
(\byear{2020})
\doiurl{10.1038/s41467-020-16863-9}
\end{barticle}
\endbibitem

%%% 38
\bibitem[\protect\citeauthoryear{Cao et~al.}{2021}]{Cao2021DiffusiveFunneling}
\begin{barticle}
\bauthor{\bsnm{Cao}, \binits{P.-C.}},
\bauthor{\bsnm{Li}, \binits{Y.}},
\bauthor{\bsnm{Peng}, \binits{Y.-G.}},
\bauthor{\bsnm{Qi}, \binits{M.}},
\bauthor{\bsnm{Huang}, \binits{W.-X.}},
\bauthor{\bsnm{Li}, \binits{P.-Q.}},
\bauthor{\bsnm{Zhu}, \binits{X.-F.}}:
\batitle{{Diffusive skin effect and topological heat funneling}}.
\bjtitle{Communications Physics}
\bvolume{4}(\bissue{1}),
\bfpage{230}
(\byear{2021})
\doiurl{10.1038/s42005-021-00731-z}
\end{barticle}
\endbibitem

%%% 39
\bibitem[\protect\citeauthoryear{Sun et~al.}{2024}]{Sun2024PhotonicEffect}
\begin{barticle}
\bauthor{\bsnm{Sun}, \binits{Y.}},
\bauthor{\bsnm{Hou}, \binits{X.}},
\bauthor{\bsnm{Wan}, \binits{T.}},
\bauthor{\bsnm{Wang}, \binits{F.}},
\bauthor{\bsnm{Zhu}, \binits{S.}},
\bauthor{\bsnm{Ruan}, \binits{Z.}},
\bauthor{\bsnm{Yang}, \binits{Z.}}:
\batitle{{Photonic Floquet Skin-Topological Effect}}.
\bjtitle{Physical Review Letters}
\bvolume{132}(\bissue{6}),
\bfpage{063804}
(\byear{2024})
\doiurl{10.1103/PhysRevLett.132.063804}
\end{barticle}
\endbibitem

%%% 40
\bibitem[\protect\citeauthoryear{Okuma et~al.}{2020}]{Okuma2020TopologicalEffects}
\begin{barticle}
\bauthor{\bsnm{Okuma}, \binits{N.}},
\bauthor{\bsnm{Kawabata}, \binits{K.}},
\bauthor{\bsnm{Shiozaki}, \binits{K.}},
\bauthor{\bsnm{Sato}, \binits{M.}}:
\batitle{{Topological Origin of Non-Hermitian Skin Effects}}.
\bjtitle{Physical Review Letters}
\bvolume{124}(\bissue{8}),
\bfpage{086801}
(\byear{2020})
\doiurl{10.1103/PhysRevLett.124.086801}
\end{barticle}
\endbibitem

%%% 41
\bibitem[\protect\citeauthoryear{Weidemann et~al.}{2020}]{Weidemann2020TopologicalLight}
\begin{barticle}
\bauthor{\bsnm{Weidemann}, \binits{S.}},
\bauthor{\bsnm{Kremer}, \binits{M.}},
\bauthor{\bsnm{Helbig}, \binits{T.}},
\bauthor{\bsnm{Hofmann}, \binits{T.}},
\bauthor{\bsnm{Stegmaier}, \binits{A.}},
\bauthor{\bsnm{Greiter}, \binits{M.}},
\bauthor{\bsnm{Thomale}, \binits{R.}},
\bauthor{\bsnm{Szameit}, \binits{A.}}:
\batitle{{Topological funneling of light}}.
\bjtitle{Science}
\bvolume{368}(\bissue{6488}),
\bfpage{311}--\blpage{314}
(\byear{2020})
\doiurl{10.1126/science.aaz8727}
\end{barticle}
\endbibitem

%%% 42
\bibitem[\protect\citeauthoryear{Leefmans et~al.}{2024}]{Leefmans2024TopologicalLaser}
\begin{barticle}
\bauthor{\bsnm{Leefmans}, \binits{C.R.}},
\bauthor{\bsnm{Parto}, \binits{M.}},
\bauthor{\bsnm{Williams}, \binits{J.}},
\bauthor{\bsnm{Li}, \binits{G.H.Y.}},
\bauthor{\bsnm{Dutt}, \binits{A.}},
\bauthor{\bsnm{Nori}, \binits{F.}},
\bauthor{\bsnm{Marandi}, \binits{A.}}:
\batitle{{Topological temporally mode-locked laser}}.
\bjtitle{Nature Physics}
\bvolume{20}(\bissue{5}),
\bfpage{852}--\blpage{858}
(\byear{2024})
\doiurl{10.1038/s41567-024-02420-4}
\end{barticle}
\endbibitem

%%% 43
\bibitem[\protect\citeauthoryear{Wang et~al.}{2023}]{Wang2023Non-HermitianMetamaterials}
\begin{botherref}
\oauthor{\bsnm{Wang}, \binits{A.}},
\oauthor{\bsnm{Meng}, \binits{Z.}},
\oauthor{\bsnm{Chen}, \binits{C.Q.}}:
{Non-Hermitian topology in static mechanical metamaterials}.
Science Advances
\textbf{9}(27)
(2023)
\doiurl{10.1126/sciadv.adf7299}
\end{botherref}
\endbibitem

%%% 44
\bibitem[\protect\citeauthoryear{Laughlin}{1983}]{Laughlin1983FQHE}
\begin{barticle}
\bauthor{\bsnm{Laughlin}, \binits{R.B.}}:
\batitle{Anomalous quantum hall effect: An incompressible quantum fluid with fractionally charged excitations}.
\bjtitle{Phys. Rev. Lett.}
\bvolume{50},
\bfpage{1395}--\blpage{1398}
(\byear{1983})
\doiurl{10.1103/PhysRevLett.50.1395}
\end{barticle}
\endbibitem

%%% 45
\bibitem[\protect\citeauthoryear{Jain}{2009}]{Jain2009BookCompositeFermions}
\begin{bbook}
\bauthor{\bsnm{Jain}, \binits{J.K.}}:
\bbtitle{Composite Fermions}.
\bpublisher{Springer},
\blocation{New York}
(\byear{2009}).
\doiurl{10.1017/CBO9780511607561} .
\burl{https://www.cambridge.org/core/books/composite-fermions/AB22E09A3F9C4E98F91E1BB447AF5778}
\end{bbook}
\endbibitem

%%% 46
\bibitem[\protect\citeauthoryear{Clark et~al.}{2020}]{Clark2020LaughlinStatesFromLight}
\begin{barticle}
\bauthor{\bsnm{Clark}, \binits{L.W.}},
\bauthor{\bsnm{Schine}, \binits{N.}},
\bauthor{\bsnm{Baum}, \binits{C.}},
\bauthor{\bsnm{Jia}, \binits{N.}},
\bauthor{\bsnm{Simon}, \binits{J.}}:
\batitle{Observation of laughlin states made of light}.
\bjtitle{Nature}
\bvolume{582}(\bissue{7810}),
\bfpage{41}--\blpage{45}
(\byear{2020})
\doiurl{10.1038/s41586-020-2318-5}
\end{barticle}
\endbibitem

%%% 47
\bibitem[\protect\citeauthoryear{Stormer et~al.}{1999}]{Stromer1999FractionalQuantumHall}
\begin{barticle}
\bauthor{\bsnm{Stormer}, \binits{H.L.}},
\bauthor{\bsnm{Tsui}, \binits{D.C.}},
\bauthor{\bsnm{Gossard}, \binits{A.C.}}:
\batitle{The fractional quantum hall effect}.
\bjtitle{Rev. Mod. Phys.}
\bvolume{71},
\bfpage{298}--\blpage{305}
(\byear{1999})
\doiurl{10.1103/RevModPhys.71.S298}
\end{barticle}
\endbibitem

%%% 48
\bibitem[\protect\citeauthoryear{Vicentini et~al.}{2018}]{Vicentini2018DrivenBoseHubbard}
\begin{barticle}
\bauthor{\bsnm{Vicentini}, \binits{F.}},
\bauthor{\bsnm{Minganti}, \binits{F.}},
\bauthor{\bsnm{Rota}, \binits{R.}},
\bauthor{\bsnm{Orso}, \binits{G.}},
\bauthor{\bsnm{Ciuti}, \binits{C.}}:
\batitle{Critical slowing down in driven-dissipative bose-hubbard lattices}.
\bjtitle{Phys. Rev. A}
\bvolume{97},
\bfpage{013853}
(\byear{2018})
\doiurl{10.1103/PhysRevA.97.013853}
\end{barticle}
\endbibitem

%%% 49
\bibitem[\protect\citeauthoryear{Carusotto and Ciuti}{2013}]{Carusotto2013QuantumFluids}
\begin{barticle}
\bauthor{\bsnm{Carusotto}, \binits{I.}},
\bauthor{\bsnm{Ciuti}, \binits{C.}}:
\batitle{Quantum fluids of light}.
\bjtitle{Rev. Mod. Phys.}
\bvolume{85},
\bfpage{299}--\blpage{366}
(\byear{2013})
\doiurl{10.1103/RevModPhys.85.299}
\end{barticle}
\endbibitem

%%% 50
\bibitem[\protect\citeauthoryear{Szameit and Rechtsman}{2024}]{Szameit2024DiscretePhotonics}
\begin{barticle}
\bauthor{\bsnm{Szameit}, \binits{A.}},
\bauthor{\bsnm{Rechtsman}, \binits{M.C.}}:
\batitle{{Discrete nonlinear topological photonics}}.
\bjtitle{Nature Physics}
\bvolume{20}(\bissue{6}),
\bfpage{905}--\blpage{912}
(\byear{2024})
\doiurl{10.1038/s41567-024-02454-8}
\end{barticle}
\endbibitem

%%% 51
\bibitem[\protect\citeauthoryear{Liu et~al.}{2021}]{Liu2021Gain-inducedInteractions}
\begin{barticle}
\bauthor{\bsnm{Liu}, \binits{Y.G.N.}},
\bauthor{\bsnm{Jung}, \binits{P.S.}},
\bauthor{\bsnm{Parto}, \binits{M.}},
\bauthor{\bsnm{Christodoulides}, \binits{D.N.}},
\bauthor{\bsnm{Khajavikhan}, \binits{M.}}:
\batitle{{Gain-induced topological response via tailored long-range interactions}}.
\bjtitle{Nature Physics 2021 17:6}
\bvolume{17}(\bissue{6}),
\bfpage{704}--\blpage{709}
(\byear{2021})
\doiurl{10.1038/s41567-021-01185-4}
\end{barticle}
\endbibitem

%%% 52
\bibitem[\protect\citeauthoryear{St-Jean et~al.}{2017}]{St-Jean2017LasingLattice}
\begin{barticle}
\bauthor{\bsnm{St-Jean}, \binits{P.}},
\bauthor{\bsnm{Goblot}, \binits{V.}},
\bauthor{\bsnm{Galopin}, \binits{E.}},
\bauthor{\bsnm{Lema{\^{i}}tre}, \binits{A.}},
\bauthor{\bsnm{Ozawa}, \binits{T.}},
\bauthor{\bsnm{Le~Gratiet}, \binits{L.}},
\bauthor{\bsnm{Sagnes}, \binits{I.}},
\bauthor{\bsnm{Bloch}, \binits{J.}},
\bauthor{\bsnm{Amo}, \binits{A.}}:
\batitle{{Lasing in topological edge states of a one-dimensional lattice}}.
\bjtitle{Nature Photonics}
\bvolume{11}(\bissue{10}),
\bfpage{651}--\blpage{656}
(\byear{2017})
\doiurl{10.1038/s41566-017-0006-2}
\end{barticle}
\endbibitem

%%% 53
\bibitem[\protect\citeauthoryear{Bahari et~al.}{2017}]{Bahari2017NonreciprocalGeometries}
\begin{barticle}
\bauthor{\bsnm{Bahari}, \binits{B.}},
\bauthor{\bsnm{Ndao}, \binits{A.}},
\bauthor{\bsnm{Vallini}, \binits{F.}},
\bauthor{\bsnm{El~Amili}, \binits{A.}},
\bauthor{\bsnm{Fainman}, \binits{Y.}},
\bauthor{\bsnm{Kant{\'{e}}}, \binits{B.}}:
\batitle{{Nonreciprocal lasing in topological cavities of arbitrary geometries}}.
\bjtitle{Science}
\bvolume{358}(\bissue{6363}),
\bfpage{636}--\blpage{640}
(\byear{2017})
\doiurl{10.1126/science.aao4551}
\end{barticle}
\endbibitem

%%% 54
\bibitem[\protect\citeauthoryear{Harari et~al.}{2018}]{Harari2018TopologicalTheory}
\begin{botherref}
\oauthor{\bsnm{Harari}, \binits{G.}},
\oauthor{\bsnm{Bandres}, \binits{M.A.}},
\oauthor{\bsnm{Lumer}, \binits{Y.}},
\oauthor{\bsnm{Rechtsman}, \binits{M.C.}},
\oauthor{\bsnm{Chong}, \binits{Y.D.}},
\oauthor{\bsnm{Khajavikhan}, \binits{M.}},
\oauthor{\bsnm{Christodoulides}, \binits{D.N.}},
\oauthor{\bsnm{Segev}, \binits{M.}}:
{Topological insulator laser: Theory}.
Science
\textbf{359}(6381)
(2018)
\doiurl{10.1126/science.aar4003}
\end{botherref}
\endbibitem

%%% 55
\bibitem[\protect\citeauthoryear{Bandres et~al.}{2018}]{Bandres2018TopologicalExperiments}
\begin{botherref}
\oauthor{\bsnm{Bandres}, \binits{M.A.}},
\oauthor{\bsnm{Wittek}, \binits{S.}},
\oauthor{\bsnm{Harari}, \binits{G.}},
\oauthor{\bsnm{Parto}, \binits{M.}},
\oauthor{\bsnm{Ren}, \binits{J.}},
\oauthor{\bsnm{Segev}, \binits{M.}},
\oauthor{\bsnm{Christodoulides}, \binits{D.N.}},
\oauthor{\bsnm{Khajavikhan}, \binits{M.}}:
{Topological insulator laser: Experiments}.
Science
\textbf{359}(6381)
(2018)
\doiurl{10.1126/science.aar4005}
\end{botherref}
\endbibitem

%%% 56
\bibitem[\protect\citeauthoryear{Amelio and Carusotto}{2020}]{Amelio2020TheoryLasers}
\begin{barticle}
\bauthor{\bsnm{Amelio}, \binits{I.}},
\bauthor{\bsnm{Carusotto}, \binits{I.}}:
\batitle{{Theory of the Coherence of Topological Lasers}}.
\bjtitle{Physical Review X}
\bvolume{10}(\bissue{4}),
\bfpage{041060}
(\byear{2020})
\doiurl{10.1103/PhysRevX.10.041060}
\end{barticle}
\endbibitem

%%% 57
\bibitem[\protect\citeauthoryear{Zeng et~al.}{2020}]{Zeng2020ElectricallyModes}
\begin{barticle}
\bauthor{\bsnm{Zeng}, \binits{Y.}},
\bauthor{\bsnm{Chattopadhyay}, \binits{U.}},
\bauthor{\bsnm{Zhu}, \binits{B.}},
\bauthor{\bsnm{Qiang}, \binits{B.}},
\bauthor{\bsnm{Li}, \binits{J.}},
\bauthor{\bsnm{Jin}, \binits{Y.}},
\bauthor{\bsnm{Li}, \binits{L.}},
\bauthor{\bsnm{Davies}, \binits{A.G.}},
\bauthor{\bsnm{Linfield}, \binits{E.H.}},
\bauthor{\bsnm{Zhang}, \binits{B.}},
\bauthor{\bsnm{Chong}, \binits{Y.}},
\bauthor{\bsnm{Wang}, \binits{Q.J.}}:
\batitle{{Electrically pumped topological laser with valley edge modes}}.
\bjtitle{Nature 2020 578:7794}
\bvolume{578}(\bissue{7794}),
\bfpage{246}--\blpage{250}
(\byear{2020})
\doiurl{10.1038/s41586-020-1981-x}
\end{barticle}
\endbibitem

%%% 58
\bibitem[\protect\citeauthoryear{Dikopoltsev et~al.}{2021}]{Dikopoltsev2021TopologicalArray}
\begin{barticle}
\bauthor{\bsnm{Dikopoltsev}, \binits{A.}},
\bauthor{\bsnm{Harder}, \binits{T.H.}},
\bauthor{\bsnm{Lustig}, \binits{E.}},
\bauthor{\bsnm{Egorov}, \binits{O.A.}},
\bauthor{\bsnm{Beierlein}, \binits{J.}},
\bauthor{\bsnm{Wolf}, \binits{A.}},
\bauthor{\bsnm{Lumer}, \binits{Y.}},
\bauthor{\bsnm{Emmerling}, \binits{M.}},
\bauthor{\bsnm{Schneider}, \binits{C.}},
\bauthor{\bsnm{H{\"{o}}fling}, \binits{S.}},
\bauthor{\bsnm{Segev}, \binits{M.}},
\bauthor{\bsnm{Klembt}, \binits{S.}}:
\batitle{{Topological insulator vertical-cavity laser array}}.
\bjtitle{Science}
\bvolume{373}(\bissue{6562}),
\bfpage{1514}--\blpage{1517}
(\byear{2021})
\doiurl{10.1126/science.abj2232}
\end{barticle}
\endbibitem

%%% 59
\bibitem[\protect\citeauthoryear{Lustig et~al.}{2018}]{Lustig2018TopologicalCrystals}
\begin{barticle}
\bauthor{\bsnm{Lustig}, \binits{E.}},
\bauthor{\bsnm{Sharabi}, \binits{Y.}},
\bauthor{\bsnm{Segev}, \binits{M.}}:
\batitle{{Topological aspects of photonic time crystals}}.
\bjtitle{Optica}
\bvolume{5}(\bissue{11}),
\bfpage{1390}
(\byear{2018})
\doiurl{10.1364/OPTICA.5.001390}
\end{barticle}
\endbibitem

%%% 60
\bibitem[\protect\citeauthoryear{Mittal et~al.}{2021}]{Mittal2021TopologicalSolitons}
\begin{barticle}
\bauthor{\bsnm{Mittal}, \binits{S.}},
\bauthor{\bsnm{Moille}, \binits{G.}},
\bauthor{\bsnm{Srinivasan}, \binits{K.}},
\bauthor{\bsnm{Chembo}, \binits{Y.K.}},
\bauthor{\bsnm{Hafezi}, \binits{M.}}:
\batitle{{Topological frequency combs and nested temporal solitons}}.
\bjtitle{Nature Physics}
\bvolume{17}(\bissue{10}),
\bfpage{1169}--\blpage{1176}
(\byear{2021})
\doiurl{10.1038/s41567-021-01302-3}
\end{barticle}
\endbibitem

%%% 61
\bibitem[\protect\citeauthoryear{Flower et~al.}{2024}]{Flower2024ObservationCombs}
\begin{barticle}
\bauthor{\bsnm{Flower}, \binits{C.J.}},
\bauthor{\bsnm{Jalali~Mehrabad}, \binits{M.}},
\bauthor{\bsnm{Xu}, \binits{L.}},
\bauthor{\bsnm{Moille}, \binits{G.}},
\bauthor{\bsnm{Suarez-Forero}, \binits{D.G.}},
\bauthor{\bsnm{{\"{O}}rsel}, \binits{O.}},
\bauthor{\bsnm{Bahl}, \binits{G.}},
\bauthor{\bsnm{Chembo}, \binits{Y.}},
\bauthor{\bsnm{Srinivasan}, \binits{K.}},
\bauthor{\bsnm{Mittal}, \binits{S.}},
\bauthor{\bsnm{Hafezi}, \binits{M.}}:
\batitle{{Observation of topological frequency combs}}.
\bjtitle{Science}
\bvolume{384}(\bissue{6702}),
\bfpage{1356}--\blpage{1361}
(\byear{2024})
\doiurl{10.1126/science.ado0053}
\end{barticle}
\endbibitem

%%% 62
\bibitem[\protect\citeauthoryear{Moussa et~al.}{2023}]{Moussa2023ObservationInterfaces}
\begin{barticle}
\bauthor{\bsnm{Moussa}, \binits{H.}},
\bauthor{\bsnm{Xu}, \binits{G.}},
\bauthor{\bsnm{Yin}, \binits{S.}},
\bauthor{\bsnm{Galiffi}, \binits{E.}},
\bauthor{\bsnm{Ra’di}, \binits{Y.}},
\bauthor{\bsnm{Al{\`{u}}}, \binits{A.}}:
\batitle{{Observation of temporal reflection and broadband frequency translation at photonic time interfaces}}.
\bjtitle{Nature Physics}
\bvolume{19}(\bissue{6}),
\bfpage{863}--\blpage{868}
(\byear{2023})
\doiurl{10.1038/s41567-023-01975-y}
\end{barticle}
\endbibitem

%%% 63
\bibitem[\protect\citeauthoryear{Mendon{\c{c}}a and Shukla}{2002}]{Mendonca2002TimeConcepts}
\begin{barticle}
\bauthor{\bsnm{Mendon{\c{c}}a}, \binits{J.T.}},
\bauthor{\bsnm{Shukla}, \binits{P.K.}}:
\batitle{{Time Refraction and Time Reflection: Two Basic Concepts}}.
\bjtitle{Physica Scripta}
\bvolume{65}(\bissue{2}),
\bfpage{160}--\blpage{163}
(\byear{2002})
\doiurl{10.1238/Physica.Regular.065a00160}
\end{barticle}
\endbibitem

%%% 64
\bibitem[\protect\citeauthoryear{Sharabi et~al.}{2022}]{Sharabi2022SpatiotemporalCrystals}
\begin{barticle}
\bauthor{\bsnm{Sharabi}, \binits{Y.}},
\bauthor{\bsnm{Dikopoltsev}, \binits{A.}},
\bauthor{\bsnm{Lustig}, \binits{E.}},
\bauthor{\bsnm{Lumer}, \binits{Y.}},
\bauthor{\bsnm{Segev}, \binits{M.}}:
\batitle{{Spatiotemporal photonic crystals}}.
\bjtitle{Optica}
\bvolume{9}(\bissue{6}),
\bfpage{585}
(\byear{2022})
\doiurl{10.1364/OPTICA.455672}
\end{barticle}
\endbibitem

%%% 65
\bibitem[\protect\citeauthoryear{Feis et~al.}{2024}]{Feis2024Spacetime-topologicalEvents}
\begin{botherref}
\oauthor{\bsnm{Feis}, \binits{J.}},
\oauthor{\bsnm{Weidemann}, \binits{S.}},
\oauthor{\bsnm{Sheppard}, \binits{T.}},
\oauthor{\bsnm{Price}, \binits{H.M.}},
\oauthor{\bsnm{Szameit}, \binits{A.}}:
{Spacetime-topological events}
(2024)
\end{botherref}
\endbibitem

%%% 66
\bibitem[\protect\citeauthoryear{Ren et~al.}{2025}]{Ren2025ObservationLattice}
\begin{barticle}
\bauthor{\bsnm{Ren}, \binits{Y.}},
\bauthor{\bsnm{Ye}, \binits{K.}},
\bauthor{\bsnm{Chen}, \binits{Q.}},
\bauthor{\bsnm{Chen}, \binits{F.}},
\bauthor{\bsnm{Zhang}, \binits{L.}},
\bauthor{\bsnm{Pan}, \binits{Y.}},
\bauthor{\bsnm{Li}, \binits{W.}},
\bauthor{\bsnm{Li}, \binits{X.}},
\bauthor{\bsnm{Zhang}, \binits{L.}},
\bauthor{\bsnm{Chen}, \binits{H.}},
\bauthor{\bsnm{Yang}, \binits{Y.}}:
\batitle{{Observation of momentum-gap topology of light at temporal interfaces in a time-synthetic lattice}}.
\bjtitle{Nature Communications 2025 16:1}
\bvolume{16}(\bissue{1}),
\bfpage{1}--\blpage{7}
(\byear{2025})
\doiurl{10.1038/s41467-025-56021-7}
\end{barticle}
\endbibitem

%%% 67
\bibitem[\protect\citeauthoryear{Shandilya et~al.}{2025}]{Shandilya2025}
\begin{barticle}
\bauthor{\bsnm{Shandilya}, \binits{P.}},
\bauthor{\bsnm{Ou}, \binits{S.-C.}},
\bauthor{\bsnm{Stone}, \binits{J.}},
\bauthor{\bsnm{Menyuk}, \binits{C.}},
\bauthor{\bsnm{Erkintalo}, \binits{M.}},
\bauthor{\bsnm{Srinivasan}, \binits{K.}},
\bauthor{\bsnm{Moille}, \binits{G.}}:
\batitle{All-optical azimuthal trapping of dissipative kerr multi-solitons for relative noise suppression}.
\bjtitle{APL Photonics}
\bvolume{10}(\bissue{1}),
\bfpage{016104}
(\byear{2025})
\doiurl{10.1063/5.0234030}
{\href{https://arxiv.org/abs/https://pubs.aip.org/aip/app/article-pdf/doi/10.1063/5.0234030/20341185/016104\_1\_5.0234030.pdf}{{https://pubs.aip.org/aip/app/article-pdf/doi/10.1063/5.0234030/20341185/016104\_1\_5.0234030.pdf}}}
\end{barticle}
\endbibitem

%%% 68
\bibitem[\protect\citeauthoryear{Wang et~al.}{2018}]{Wang2018}
\begin{barticle}
\bauthor{\bsnm{Wang}, \binits{S.}},
\bauthor{\bsnm{Carruthers}, \binits{T.F.}},
\bauthor{\bsnm{Menyuk}, \binits{C.R.}}:
\batitle{Efficiently modeling the noise performance of short-pulse lasers with a computational implementation of dynamical methods}.
\bjtitle{J. Opt. Soc. Am. B}
\bvolume{35}(\bissue{10}),
\bfpage{2521}--\blpage{2531}
(\byear{2018})
\doiurl{10.1364/JOSAB.35.002521}
\end{barticle}
\endbibitem

%%% 69
\bibitem[\protect\citeauthoryear{Lado and Zilberberg}{2019}]{Lado2019HarperHeisenberg}
\begin{barticle}
\bauthor{\bsnm{Lado}, \binits{J.L.}},
\bauthor{\bsnm{Zilberberg}, \binits{O.}}:
\batitle{Topological spin excitations in harper-heisenberg spin chains}.
\bjtitle{Phys. Rev. Res.}
\bvolume{1},
\bfpage{033009}
(\byear{2019})
\doiurl{10.1103/PhysRevResearch.1.033009}
\end{barticle}
\endbibitem

%%% 70
\bibitem[\protect\citeauthoryear{Lustig et~al.}{2023}]{Lustig2023Time-refractionModulation}
\begin{barticle}
\bauthor{\bsnm{Lustig}, \binits{E.}},
\bauthor{\bsnm{Segal}, \binits{O.}},
\bauthor{\bsnm{Saha}, \binits{S.}},
\bauthor{\bsnm{Bordo}, \binits{E.}},
\bauthor{\bsnm{Chowdhury}, \binits{S.N.}},
\bauthor{\bsnm{Sharabi}, \binits{Y.}},
\bauthor{\bsnm{Fleischer}, \binits{A.}},
\bauthor{\bsnm{Boltasseva}, \binits{A.}},
\bauthor{\bsnm{Cohen}, \binits{O.}},
\bauthor{\bsnm{Shalaev}, \binits{V.M.}},
\bauthor{\bsnm{Segev}, \binits{M.}}:
\batitle{{Time-refraction optics with single cycle modulation}}.
\bjtitle{Nanophotonics}
\bvolume{12}(\bissue{12}),
\bfpage{2221}--\blpage{2230}
(\byear{2023})
\doiurl{10.1515/nanoph-2023-0126}
\end{barticle}
\endbibitem

%%% 71
\bibitem[\protect\citeauthoryear{G{\"{u}}nter et~al.}{2009}]{Gunter2009Sub-cycleInteraction}
\begin{barticle}
\bauthor{\bsnm{G{\"{u}}nter}, \binits{G.}},
\bauthor{\bsnm{Anappara}, \binits{A.A.}},
\bauthor{\bsnm{Hees}, \binits{J.}},
\bauthor{\bsnm{Sell}, \binits{A.}},
\bauthor{\bsnm{Biasiol}, \binits{G.}},
\bauthor{\bsnm{Sorba}, \binits{L.}},
\bauthor{\bsnm{De~Liberato}, \binits{S.}},
\bauthor{\bsnm{Ciuti}, \binits{C.}},
\bauthor{\bsnm{Tredicucci}, \binits{A.}},
\bauthor{\bsnm{Leitenstorfer}, \binits{A.}},
\bauthor{\bsnm{Huber}, \binits{R.}}:
\batitle{{Sub-cycle switch-on of ultrastrong light–matter interaction}}.
\bjtitle{Nature}
\bvolume{458}(\bissue{7235}),
\bfpage{178}--\blpage{181}
(\byear{2009})
\doiurl{10.1038/nature07838}
\end{barticle}
\endbibitem

%%% 72
\bibitem[\protect\citeauthoryear{Jackiw}{1997}]{Jackiw1997CurrentSoliton}
\begin{barticle}
\bauthor{\bsnm{Jackiw}, \binits{R.}}:
\batitle{A nonrelativistic chiral soliton in one dimension}.
\bjtitle{Journal of Nonlinear Mathematical Physics}
\bvolume{4}(\bissue{3–4}),
\bfpage{261}
(\byear{1997})
\doiurl{10.2991/jnmp.1997.4.3-4.2}
\end{barticle}
\endbibitem

%%% 73
\bibitem[\protect\citeauthoryear{Dutta~Roy et~al.}{2019}]{TriangularWormholes2019Sayan}
\begin{barticle}
\bauthor{\bsnm{Dutta~Roy}, \binits{P.}},
\bauthor{\bsnm{Das}, \binits{J.}},
\bauthor{\bsnm{Kar}, \binits{S.}}:
\batitle{Quasi-normal modes in a symmetric triangular barrier}.
\bjtitle{The European Physical Journal Plus}
\bvolume{134}(\bissue{11}),
\bfpage{571}
(\byear{2019})
\doiurl{10.1140/epjp/i2019-12928-y}
\end{barticle}
\endbibitem

%%% 74
\bibitem[\protect\citeauthoryear{Khurgin et~al.}{2014}]{Khurgin2014}
\begin{barticle}
\bauthor{\bsnm{Khurgin}, \binits{J.B.}},
\bauthor{\bsnm{Dikmelik}, \binits{Y.}},
\bauthor{\bsnm{Hugi}, \binits{A.}},
\bauthor{\bsnm{Faist}, \binits{J.}}:
\batitle{{Coherent frequency combs produced by self frequency modulation in quantum cascade lasers}}.
\bjtitle{Applied Physics Letters}
\bvolume{104}(\bissue{8}),
\bfpage{081118}
(\byear{2014})
\doiurl{10.1063/1.4866868}
\end{barticle}
\endbibitem

%%% 75
\bibitem[\protect\citeauthoryear{Opa{\v{c}}ak and Schwarz}{2019}]{Opacak2019b}
\begin{barticle}
\bauthor{\bsnm{Opa{\v{c}}ak}, \binits{N.}},
\bauthor{\bsnm{Schwarz}, \binits{B.}}:
\batitle{{Theory of Frequency-Modulated Combs in Lasers with Spatial Hole Burning, Dispersion, and Kerr Nonlinearity}}.
\bjtitle{Physical Review Letters}
\bvolume{123}(\bissue{24}),
\bfpage{243902}
(\byear{2019})
\doiurl{10.1103/PhysRevLett.123.243902}
\end{barticle}
\endbibitem

%%% 76
\bibitem[\protect\citeauthoryear{Burghoff}{2020}]{Burghoff2020d}
\begin{barticle}
\bauthor{\bsnm{Burghoff}, \binits{D.}}:
\batitle{{Unraveling the origin of frequency modulated combs using active cavity mean-field theory}}.
\bjtitle{Optica}
\bvolume{7}(\bissue{12}),
\bfpage{1781}
(\byear{2020})
\doiurl{10.1364/optica.408917}
\end{barticle}
\endbibitem

%%% 77
\bibitem[\protect\citeauthoryear{Heckelmann et~al.}{2023}]{Heckelmann2023QuantumLaser}
\begin{barticle}
\bauthor{\bsnm{Heckelmann}, \binits{I.}},
\bauthor{\bsnm{Bertrand}, \binits{M.}},
\bauthor{\bsnm{Dikopoltsev}, \binits{A.}},
\bauthor{\bsnm{Beck}, \binits{M.}},
\bauthor{\bsnm{Scalari}, \binits{G.}},
\bauthor{\bsnm{Faist}, \binits{J.}}:
\batitle{{Quantum walk comb in a fast gain laser}}.
\bjtitle{Science}
\bvolume{282}(\bissue{6669}),
\bfpage{434}--\blpage{438}
(\byear{2023})
\doiurl{10.1126/SCIENCE.ADJ3858}
\end{barticle}
\endbibitem

%%% 78
\bibitem[\protect\citeauthoryear{T{\"{a}}schler et~al.}{2023}]{Taschler2023h}
\begin{barticle}
\bauthor{\bsnm{T{\"{a}}schler}, \binits{P.}},
\bauthor{\bsnm{Forrer}, \binits{A.}},
\bauthor{\bsnm{Bertrand}, \binits{M.}},
\bauthor{\bsnm{Kapsalidis}, \binits{F.}},
\bauthor{\bsnm{Beck}, \binits{M.}},
\bauthor{\bsnm{Faist}, \binits{J.}}:
\batitle{{Asynchronous Upconversion Sampling of Frequency Modulated Combs}}.
\bjtitle{Laser {\&} Photonics Reviews}
\bvolume{17}(\bissue{4}),
\bfpage{2200590}
(\byear{2023})
\doiurl{10.1002/lpor.202200590}
\end{barticle}
\endbibitem

%%% 79
\bibitem[\protect\citeauthoryear{Dikopoltsev et~al.}{2025a}]{Dikopoltsev2025TheLaser}
\begin{barticle}
\bauthor{\bsnm{Dikopoltsev}, \binits{A.}},
\bauthor{\bsnm{Heckelmann}, \binits{I.}},
\bauthor{\bsnm{Schneider}, \binits{B.}},
\bauthor{\bsnm{Bertrand}, \binits{M.}},
\bauthor{\bsnm{Faist}, \binits{J.}}:
\batitle{{The theory of the quantum walk comb laser}}.
\bjtitle{Nanophotonics}
(\byear{2025})
\doiurl{10.1515/nanoph-2024-0768}
\end{barticle}
\endbibitem

%%% 80
\bibitem[\protect\citeauthoryear{Dikopoltsev et~al.}{2025b}]{Dikopoltsev2025CollectiveDimensions}
\begin{botherref}
\oauthor{\bsnm{Dikopoltsev}, \binits{A.}},
\oauthor{\bsnm{Heckelmann}, \binits{I.}},
\oauthor{\bsnm{Bertrand}, \binits{M.}},
\oauthor{\bsnm{Beck}, \binits{M.}},
\oauthor{\bsnm{Scalari}, \binits{G.}},
\oauthor{\bsnm{Zilberberg}, \binits{O.}},
\oauthor{\bsnm{Faist}, \binits{J.}}:
{Collective quench dynamics of active photonic lattices in synthetic dimensions}.
Nature Physics 2025,
1--7
(2025)
\doiurl{10.1038/s41567-025-02880-2}
\end{botherref}
\endbibitem

%%% 81
\bibitem[\protect\citeauthoryear{Humbard and Burghoff}{2022}]{Humbard2022AnalyticalStates}
\begin{barticle}
\bauthor{\bsnm{Humbard}, \binits{L.}},
\bauthor{\bsnm{Burghoff}, \binits{D.}}:
\batitle{{Analytical theory of frequency-modulated combs: generalized mean-field theory, complex cavities, and harmonic states}}.
\bjtitle{Optics Express}
\bvolume{30}(\bissue{4}),
\bfpage{5376}
(\byear{2022})
\doiurl{10.1364/OE.445570}
\end{barticle}
\endbibitem

%%% 82
\bibitem[\protect\citeauthoryear{Chiocchetta and Carusotto}{2013}]{Chiocchetta2013Non-equilibriumDimensions}
\begin{barticle}
\bauthor{\bsnm{Chiocchetta}, \binits{A.}},
\bauthor{\bsnm{Carusotto}, \binits{I.}}:
\batitle{{Non-equilibrium quasi-condensates in reduced dimensions}}.
\bjtitle{Europhysics Letters}
\bvolume{102}(\bissue{6}),
\bfpage{67007}
(\byear{2013})
\doiurl{10.1209/0295-5075/102/67007}
\end{barticle}
\endbibitem

%%% 83
\bibitem[\protect\citeauthoryear{Faist et~al.}{1994}]{Faist1994}
\begin{barticle}
\bauthor{\bsnm{Faist}, \binits{J.}},
\bauthor{\bsnm{Capasso}, \binits{F.}},
\bauthor{\bsnm{Sivco}, \binits{D.L.}},
\bauthor{\bsnm{Sirtori}, \binits{C.}},
\bauthor{\bsnm{Hutchinson}, \binits{A.L.}},
\bauthor{\bsnm{Cho}, \binits{A.Y.}}:
\batitle{{Quantum Cascade Laser}}.
\bjtitle{Science}
\bvolume{264}(\bissue{5158}),
\bfpage{553}--\blpage{556}
(\byear{1994})
\doiurl{10.1126/science.264.5158.553}
\end{barticle}
\endbibitem

%%% 84
\bibitem[\protect\citeauthoryear{Han et~al.}{2020}]{Han2020a}
\begin{barticle}
\bauthor{\bsnm{Han}, \binits{Z.}},
\bauthor{\bsnm{Ren}, \binits{D.}},
\bauthor{\bsnm{Burghoff}, \binits{D.}}:
\batitle{{Sensitivity of SWIFT spectroscopy}}.
\bjtitle{Optics Express}
\bvolume{28}(\bissue{5}),
\bfpage{6002}
(\byear{2020})
\doiurl{10.1364/OE.382243}
\end{barticle}
\endbibitem

%%% 85
\bibitem[\protect\citeauthoryear{Schneider et~al.}{2021}]{Schneider2021c}
\begin{barticle}
\bauthor{\bsnm{Schneider}, \binits{B.}},
\bauthor{\bsnm{Kapsalidis}, \binits{F.}},
\bauthor{\bsnm{Bertrand}, \binits{M.}},
\bauthor{\bsnm{Singleton}, \binits{M.}},
\bauthor{\bsnm{Hillbrand}, \binits{J.}},
\bauthor{\bsnm{Beck}, \binits{M.}},
\bauthor{\bsnm{Faist}, \binits{J.}}:
\batitle{{Controlling Quantum Cascade Laser Optical Frequency Combs through Microwave Injection}}.
\bjtitle{Laser {\&} Photonics Reviews}
\bvolume{15}(\bissue{12}),
\bfpage{2100242}
(\byear{2021})
\doiurl{10.1002/lpor.202100242}
\end{barticle}
\endbibitem

%%% 86
\bibitem[\protect\citeauthoryear{Singleton et~al.}{2018}]{Singleton2018a}
\begin{barticle}
\bauthor{\bsnm{Singleton}, \binits{M.}},
\bauthor{\bsnm{Jouy}, \binits{P.}},
\bauthor{\bsnm{Beck}, \binits{M.}},
\bauthor{\bsnm{Faist}, \binits{J.}}:
\batitle{{Evidence of linear chirp in mid-infrared quantum cascade lasers}}.
\bjtitle{Optica}
\bvolume{5}(\bissue{8}),
\bfpage{948}
(\byear{2018})
\doiurl{10.1364/optica.5.000948}
\end{barticle}
\endbibitem

%%% 87
\bibitem[\protect\citeauthoryear{Hillbrand et~al.}{2019}]{Hillbrand2019}
\begin{barticle}
\bauthor{\bsnm{Hillbrand}, \binits{J.}},
\bauthor{\bsnm{Andrews}, \binits{A.M.}},
\bauthor{\bsnm{Detz}, \binits{H.}},
\bauthor{\bsnm{Strasser}, \binits{G.}},
\bauthor{\bsnm{Schwarz}, \binits{B.}}:
\batitle{{Coherent injection locking of quantum cascade laser frequency combs}}.
\bjtitle{Nature Photonics}
\bvolume{13}(\bissue{2}),
\bfpage{101}--\blpage{104}
(\byear{2019})
\doiurl{10.1038/s41566-018-0320-3}
\end{barticle}
\endbibitem

%%% 88
\bibitem[\protect\citeauthoryear{Harder et~al.}{1983}]{Harder1983a}
\begin{barticle}
\bauthor{\bsnm{Harder}, \binits{C.}},
\bauthor{\bsnm{Vahala}, \binits{K.}},
\bauthor{\bsnm{Yariv}, \binits{A.}}:
\batitle{{Measurement of the linewidth enhancement factor {$\alpha$} of semiconductor lasers}}.
\bjtitle{Applied Physics Letters}
\bvolume{42}(\bissue{4}),
\bfpage{328}--\blpage{330}
(\byear{1983})
\doiurl{10.1063/1.93921}
\end{barticle}
\endbibitem

%%% 89
\bibitem[\protect\citeauthoryear{Aellen et~al.}{2006}]{Aellen2006a}
\begin{barticle}
\bauthor{\bsnm{Aellen}, \binits{T.}},
\bauthor{\bsnm{Maulini}, \binits{R.}},
\bauthor{\bsnm{Terazzi}, \binits{R.}},
\bauthor{\bsnm{Hoyler}, \binits{N.}},
\bauthor{\bsnm{Giovannini}, \binits{M.}},
\bauthor{\bsnm{Faist}, \binits{J.}},
\bauthor{\bsnm{Blaser}, \binits{S.}},
\bauthor{\bsnm{Hvozdara}, \binits{L.}}:
\batitle{{Direct measurement of the linewidth enhancement factor by optical heterodyning of an amplitude-modulated quantum cascade laser}}.
\bjtitle{Applied Physics Letters}
\bvolume{89}(\bissue{9}),
\bfpage{091121}
(\byear{2006})
\doiurl{10.1063/1.2345035}
\end{barticle}
\endbibitem

%%% 90
\bibitem[\protect\citeauthoryear{Opa{\v{c}}ak et~al.}{2021}]{Opacak2021SpectrallyComb}
\begin{barticle}
\bauthor{\bsnm{Opa{\v{c}}ak}, \binits{N.}},
\bauthor{\bsnm{Opa{\v{c}}ak}, \binits{N.}},
\bauthor{\bsnm{Pilat}, \binits{F.}},
\bauthor{\bsnm{Kazakov}, \binits{D.}},
\bauthor{\bsnm{Kazakov}, \binits{D.}},
\bauthor{\bsnm{Cin}, \binits{S.D.}},
\bauthor{\bsnm{Ramer}, \binits{G.}},
\bauthor{\bsnm{Lendl}, \binits{B.}},
\bauthor{\bsnm{Capasso}, \binits{F.}},
\bauthor{\bsnm{Schwarz}, \binits{B.}},
\bauthor{\bsnm{Schwarz}, \binits{B.}}:
\batitle{{Spectrally resolved linewidth enhancement factor of a semiconductor frequency comb}}.
\bjtitle{Optica, Vol. 8, Issue 9, pp. 1227-1230}
\bvolume{8}(\bissue{9}),
\bfpage{1227}--\blpage{1230}
(\byear{2021})
\doiurl{10.1364/OPTICA.428096}
\end{barticle}
\endbibitem

%%% 91
\bibitem[\protect\citeauthoryear{Kapsalidis et~al.}{2021}]{Kapsalidis2021e}
\begin{barticle}
\bauthor{\bsnm{Kapsalidis}, \binits{F.}},
\bauthor{\bsnm{Schneider}, \binits{B.}},
\bauthor{\bsnm{Singleton}, \binits{M.}},
\bauthor{\bsnm{Bertrand}, \binits{M.}},
\bauthor{\bsnm{Gini}, \binits{E.}},
\bauthor{\bsnm{Beck}, \binits{M.}},
\bauthor{\bsnm{Faist}, \binits{J.}}:
\batitle{{Mid-infrared quantum cascade laser frequency combs with a microstrip-like line waveguide geometry}}.
\bjtitle{Applied Physics Letters}
\bvolume{118}(\bissue{7}),
\bfpage{071101}
(\byear{2021})
\doiurl{10.1063/5.0040882}
\end{barticle}
\endbibitem

%%% 92
\bibitem[\protect\citeauthoryear{wen Xiao}{2009}]{Xiao2009TransformationTheory}
\begin{botherref}
\oauthor{\bsnm{Xiao}, \binits{M.-w.}}:
Theory of transformation for the diagonalization of quadratic Hamiltonians
(2009).
\url{https://arxiv.org/abs/0908.0787}
\end{botherref}
\endbibitem

\end{thebibliography}
\end{document}